\documentclass[a4paper, USenglish, numberwithinsect, cleveref, pdfa]{lipics-v2021}

\usepackage{hyperref}

\pdfoutput=1

\hideLIPIcs

\nolinenumbers

\title{Strategic Dominance:\\A New Preorder for Nondeterministic Processes}
\titlerunning{Strategic Dominance}

\author{Thomas A. {Henzinger}}{Institute of Science and Technology Austria (ISTA), Klosterneuburg, Austria \and \url{https://pub.ista.ac.at/~tah/}}{tah@ist.ac.at}{https://orcid.org/0000-0002-2985-7724}{}
\author{Nicolas Mazzocchi}{Slovak University of Technology in Bratislava, Slovak Republic \and Institute of Science and Technology Austria (ISTA), Klosterneuburg, Austria \and \url{https://mazzocchi.github.io/}}{nicolas.mazzocchi@stuba.sk}{https://orcid.org/0000-0001-6425-5369}{}
\author{N. Ege {Sara\c{c}}}{Institute of Science and Technology Austria (ISTA), Klosterneuburg, Austria \and \url{https://egesarac.github.io/}}{esarac@ist.ac.at}{https://orcid.org/0009-0000-2866-8078}{}

\authorrunning{T.\,A. Henzinger and N. Mazzocchi and N.\,E. Sara\c{c}}

\Copyright{Thomas A. Henzinger and Nicolas Mazzocchi and N. Ege {Sara\c{c}}}



\funding{This work was supported in part by the ERC-2020-AdG 101020093. N.~Mazzocchi was affiliated with ISTA when this work was submitted for publication.}

\acknowledgements{We thank the anonymous reviewers for their helpful comments.}

\EventEditors{Rupak Majumdar and Alexandra Silva}
\EventNoEds{2}
\EventLongTitle{35th International Conference on Concurrency Theory (CONCUR 2024)}
\EventShortTitle{CONCUR 2024}
\EventAcronym{CONCUR}
\EventYear{2024}
\EventDate{September 9--13, 2024}
\EventLocation{Calgary, Canada}
\EventLogo{}
\SeriesVolume{311}
\ArticleNo{8}

\usepackage{amssymb}
\usepackage{amsmath}
\usepackage{mathtools}
\usepackage{wasysym}
\usepackage{xspace}
\usepackage{colortbl}
\usepackage{cleveref}
\usepackage{multicol}
\usepackage{enumitem}

\usepackage{todonotes}

\usepackage{tikz}
\usetikzlibrary{backgrounds, calc}
\tikzstyle{state}=[thick,minimum size=18pt, circle,draw]
\tikzstyle{transition}=[->,thick,>=stealth,shorten >=1pt,shorten <=1pt, font=\small]
\tikzstyle{initial}=[after node path={%
	[to path={[transition] (\tikztostart) -- (\tikztotarget)}]%
	(\tikzlastnode.180)++(180:13pt) edge (\tikzlastnode)%
}]
\tikzstyle{final}=[after node path={ node[state, scale=.8] at (\tikzlastnode) {} }]
\tikzset{
	bg/.default={},
	bg/.style={execute at end picture={
			\begin{scope}[on background layer]
				\node[xshift=-1mm, yshift=-1mm] (sw) at (current bounding box.south west) {};
				\node[xshift=1mm, yshift=1mm] (ne) at (current bounding box.north east) {};
				\node[xshift=1mm, yshift=-1mm] (nw) at (current bounding box.north west) {};
				\fill[fill=black!10,rounded corners] (sw) rectangle (ne);
				
				\ifx&#1&\else
				\node[anchor=north east, xshift=2pt] at (nw) {#1};
				\fi
			\end{scope}
	}},
}
\newcommand{\picturescaletikz}{.92}

\newcommand{\NN}{\ensuremath{\mathbb{N}}\xspace}
\newcommand{\QQ}{\ensuremath{\mathbb{Q}}\xspace}
\newcommand{\RR}{\ensuremath{\mathbb{R}}\xspace}

\newcommand{\A}{\ensuremath{\mathcal{A}}\xspace}
\newcommand{\B}{\ensuremath{\mathcal{B}}\xspace}
\newcommand{\C}{\ensuremath{\mathcal{C}}\xspace}

\newcommand{\T}{\ensuremath{\mathcal{T}}\xspace}

\newcommand{\printValueFunction}[1]{\mathsf{#1}}

\newcommand{\Inf}{\printValueFunction{Inf}}
\newcommand{\Sup}{\printValueFunction{Sup}}

\newcommand{\LimInf}{\printValueFunction{LimInf}}
\newcommand{\LimSup}{\printValueFunction{LimSup}}

\newcommand{\weakcommitsim}{\mathrel{\raisebox{0pt}[6pt][2pt]{\makebox[\widthof{${\subseteq}$}][c]{\makebox[0pt]{\:\:\,\rotatebox[origin=c]{90}{\raisebox{1pt}{$-$}}}\makebox[0pt]{$\subseteq$}}}}}

\newcommand{\weakblindsim}{\mathrel{\raisebox{0pt}[6pt][2pt]{\makebox[\widthof{${\supseteq}$}][c]{\makebox[0pt][c]{\scalebox{1.4}{$\bullet$}~}\makebox[0pt][c]{\scalebox{.6}{\raisebox{1.2pt}{$\blacksquare$}}}\makebox[0pt][c]{\scalebox{.6}{\raisebox{3.35pt}{$\blacksquare$}}}\makebox[0pt][c]{\raisebox{-3.8pt}{$-$}\,}}}}}

\newcommand{\commitsim}{\mathrel{\trianglelefteq}}

\newcommand{\blindsim}{\mathrel{\makebox[\widthof{$\blacktriangleleft$}][c]{\makebox[0pt]{$\leq$}\makebox[0pt]{\raisebox{1pt}{$\blacktriangleleft$}}}}}

\newcommand{\simulation}{\mathrel{\preceq}}

\newcommand{\inclusion}{\mathrel{\subseteq}}

\renewcommand{\st}{\;\ifnum\currentgrouptype=16 \middle\fi|\;}
\newcommand{\entry}[2][!]{\ifx!#1\begin{tabular}{p{\linewidth}}#2\end{tabular}\else\begin{tabular}{p{#1}}#2\end{tabular}\fi}

\newcommand{\printComplexityClass}[1]{{\textsc{#1}}\xspace}

\newcommand{\PTime}{\printComplexityClass{PTime}}
\newcommand{\NPTime}{\printComplexityClass{NP}}
\newcommand{\coNPTime}{\textmd{co-}\printComplexityClass{NP}}
\newcommand{\PSpace}{\printComplexityClass{PSpace}}
\newcommand{\PSpaceC}{\printComplexityClass{PSpace}\textmd{-complete}\xspace}
\newcommand{\PSpaceH}{\printComplexityClass{PSpace}\textmd{-hard}\xspace}
\newcommand{\ExpTime}{\printComplexityClass{ExpTime}}

\def\fprime{f_{{\it prime}}}
\def\gprime{g_{{\it prime}}}

\begin{document}
	\maketitle
	
	\begin{abstract}
	We study the following refinement relation between nondeterministic state-transition models:
	model $\B$ {\em strategically dominates} model $\A$ iff
	every deterministic refinement of $\A$ is language contained in some deterministic refinement of~$\B$.
	While language containment is trace inclusion,
	and the (fair) simulation preorder coincides with tree inclusion, 
	strategic dominance falls strictly between the two and can be characterized as ``strategy inclusion''
	between $\A$ and~$\B$:
	every strategy that resolves the nondeterminism of $\A$ is dominated by a strategy that resolves the
	nondeterminism of~$\B$.
	Strategic dominance can be checked in 2-\ExpTime by a decidable first-order Presburger logic with
	quantification over words and strategies, called {\em resolver logic}.
	We give several other applications of resolver logic,
	including checking the co-safety, co-liveness, and history-determinism of boolean and quantitative automata,
	and checking the inclusion between hyperproperties that are specified by nondeterministic boolean and
	quantitative automata.		
	\end{abstract}
	
	\section{Introduction}
	Nondeterminism is a powerful mechanism for varying the degree of detail shown in a state-transition model
	of a system.
	Intuitively, a nondeterministic model captures the set of possible deterministic implementations.
	Consider the process model $ab+ac$.
	This model allows two possible deterministic implementations, $ab$ and~$ac$.
	Mathematically, each deterministic implementation of a nondeterministic model corresponds to a
	{\em strategy\/} for resolving the nondeterminism,
	i.e., a function $f$ that maps a finite run $h$ through the model (the ``history'') and a new letter
	$\sigma$ to a successor state $f(h,\sigma)$ allowed by the model.
	Consider the recursive process model $X=abX+acX$,
	which repeatedly chooses either the $ab$ branch or the $ac$ branch.
	There are infinitely many different strategies to resolve the repeated nondeterminism,
	e.g., the strategy $\fprime$ whose $n$th choice is the $ab$ branch if $n=1$ or $n$ is prime,
	and whose $n$th choice is the $ac$ branch when $n>1$ and $n$ is composite.
	
	Two models that describe the same system at different levels of detail are related by a preorder.
	Milner's simulation preorder and the trace-inclusion preorder represent two particularly paradigmatic and
	widely studied examples of preorders on state-transition models,
	but many variations and other refinement preorders can be found in the literature
	\cite{DBLP:books/el/01/Glabbeek01,DBLP:conf/concur/Glabbeek93,DBLP:books/daglib/0096285,DBLP:journals/tcs/Brookes07,DBLP:journals/tcs/OHearn07}.
	In its purest form,
	the {\em linear-time} view of model refinement postulates that for two models $\A$ and~$\B$,
	for $\B$ to describe the same system as $\A$ at a higher level of abstraction, 
	the less precise model $\B$ must allow all traces that are possible in the more refined model~$\A$
	while potentially allowing more traces.
	In contrast,
	the pure {\em branching-time} view of model refinement postulates that for two models $\A$ and~$\B$,
	for $\B$ to describe the same system as $\A$ at a higher level of abstraction, 
	the less precise model $\B$ must allow all subtrees of the full computation tree of the more refined
	model~$\A$ (modulo its isomorphic subtrees) while potentially allowing more trees.
	This tree-inclusion view of branching time corresponds to the simulation preorder and its
	generalization to fair simulation~\cite{DBLP:journals/iandc/HenzingerKR02}.
	
	In this paper, we introduce and study a third fundamental view of model
	refinement---{\em strategy inclusion}---which lies strictly between trace inclusion and tree inclusion.
	According to the strategic view,
	for $\B$ to describe the same system as $\A$ at a higher level of abstraction,
	the less precise model $\B$ must allow all deterministic implementations that are possible for the more
	refined model~$\A$
	(but $\B$ may allow strictly more implementations than~$\A$).
	Mathematically, we say that $\B$ ({\em strategically}) {\em dominates} $\A$ if
	for every strategy $f$ resolving the nondeterminism of~$\A$,
	there exists a strategy $g$ for resolving the nondeterminism of $\B$
	such that every trace of $\A^f$
	(the deterministic result of applying the strategy $f$ to the nondeterministic model~$\A$)
	corresponds to a trace of~$\B^g$.
	
	Strategic dominance, like simulation, is a branching-time preorder,
	but while simulation corresponds to inclusion of all subtrees of the full computation tree of a model, dominance corresponds to the inclusion of all {\em deterministic} subtrees.
	To see that dominance is strictly coarser than simulation,
	recall the model $X=abX+acX$ and compare it with the model
	$Y\ =\ ababY + abacY + acabY + acacY$.
	Both models $X$ and $Y$ have the same traces, but $Y$ does not simulate~$X$.
	Nonetheless, for every strategy resolving the nondeterminism of $X$ there exists a strategy resolving
	the nondeterminism of $Y$ that produces same infinite trace (and vice versa).
	In particular, the strategy $\gprime$ for $Y$ that dominates the strategy $\fprime$ for $X$ makes the
	following choices:
	initially $\gprime$ chooses the $abab$ branch (since 2 is prime) and thereafter, for all $n>1$,
	the $n$th choice of $\gprime$ is $abac$ if $2n+1$ is prime,
	and $acac$ otherwise (because $2n+2$ is never prime).
	To see that strategic dominance is strictly finer than trace inclusion,
	the standard example of comparing $a(b+c)$ with $ab+ac$ will do;
	these two process models are trace equivalent,
	but only the former has a deterministic implementation that includes both traces.

	\smallskip\noindent
	{\color{lipicsGray}\bfseries Relation to prior work.~}
	Our motivation for studying the strategic view of model refinement originated from the definition of
	{\em history-determinism}
	\cite{DBLP:conf/csl/HenzingerP06,DBLP:journals/siglog/BokerL23}.
	A state-transition model $\A$ is history-deterministic if there exists a strategy for resolving the
	nondeterminism of $\A$ which captures exactly the language of~$\A$.
	In other words, the nondeterminism of a history-deterministic model can be resolved on-the-fly,
	by looking only at the history, without making guesses about the future.
	
	Strategies (a.k.a.\ policies) are a central concept of game theory,
	and our resolvers correspond mathematically to deterministic strategies of games played on graphs.
	However, we study the {\em single-player}
	case---where the player resolves the nondeterminism of a state-transition model---and
	not the multi-player case that arises in game models
	\cite{DBLP:conf/concur/AlurHKV98,DBLP:journals/iandc/ChatterjeeHP10}.
	Besides general strategies, we consider the special cases of {\em finite-state} strategies
	(which can remember only a finite number of bits about the history),
	and of {\em positional} strategies (which have no memory about the history).
	Positional (a.k.a.\ memoryless) strategies correspond, in our setting, to the removal of nondeterministic
	edges from a model (``edge pruning'').
	To the best of our knowledge, strategic dominance is a novel relation that has not been studied before.
	In particular, the seminal works on the linear time-branching time spectrum of sequential processes \cite{DBLP:books/el/01/Glabbeek01,DBLP:conf/concur/Glabbeek93} do not present a comparable relation.
	In general, these works do not involve a game-theoretic view of nondeterminism, which is a crucial aspect captured through our use of resolvers.
	In this way, the game-theoretic view adds a new dimension to the spectrum of process preorders.	
	
	For generality,
	we study strategies for state-transition models in a {\em quantitative} setting,
	where all states have outgoing transitions and all transitions have numeric weights
	\cite{DBLP:journals/tocl/ChatterjeeDH10}.
	In this setting, every finite path through a state-transition model can be extended,
	every infinite path is assigned a numeric value,
	and the values of different paths can be compared.
	The trace preorder between $\A$ and $\B$ requires that for every infinite word~$w$,
	for every run of $\A$ on $w$ there exists a run of $\B$ on $w$ of equal or greater value.
	All refinement relations we consider are refinements of the trace preorder.
	The quantitative setting generalizes the boolean setting
	(take $\{0,1\}$ as the value set, and assign the value~1 to a run iff the run is accepted by the model),
	and it generalizes many common acceptance conditions on finite and infinite runs
	($\Sup$ and $\Inf$ values correspond to reachability and safety acceptance;
	$\LimSup$ and $\LimInf$ values to B\"uchi and coB\"uchi acceptance).
	Extensions of simulation and inclusion preorders to this setting have been studied in \cite{DBLP:journals/tocl/ChatterjeeDH10}.

	\smallskip\noindent
	{\color{lipicsGray}\bfseries Contributions of this paper.~}
	Our results are threefold.
	First, we define {\em resolver logic} as a first-order logic that is interpreted over a set $\mathsf{QA}$ of
	quantitative finite automata over infinite words.
	Resolver logic quantifies over infinite words~$w$, strategies $f$ that resolve the nondeterminism of
	automata $\A \in \mathsf{QA}$ (so-called ``resolvers''), and natural numbers.
	The existentially quantified formulas are built from terms of the form $\A^f(w)$ using Presburger arithmetic.
	The term $\A^f(w)$ denotes the value of the unique run of the automaton $\A$ over the word $w$ when all
	nondeterministic choices are resolved by the strategy~$f$.
	We show that model-checking problem for resolver logic---i.e., the problem of deciding if a closed formula
	$\varphi$ is true over a given set ${\cal A}$ of automata---can be solved in $d$-\ExpTime,
	where $d$ is the number of quantifier switches in~$\varphi$.
	Our model-checking algorithm uses automata-theoretic constructions over parity tree automata that
	represent resolvers.
	A main difference of resolver logic to strategy logics \cite{DBLP:journals/iandc/ChatterjeeHP10,DBLP:conf/fsttcs/MogaveroMV10}, besides the handling of quantitative constraints and Presburger arithmetic, lies in the quantification over infinite words.
	This quantification is not present in strategy logics, and lets us define strategic dominance and other inclusion-based relations.
	
	Second, we define in resolver logic {\em eight different resolver relations} between quantitative automata,
	including the strategic-dominance relation explained above.
	The eight relations are obtained from each other by reordering quantifiers in resolver logic.
	We show that six of the eight relations are preorders:
	one coincides with simulation,
	four coincide with trace inclusion,
	and the sixth---strategic dominance---lies strictly between simulation and trace inclusion.
	The remaining two relations are finer than simulation and transitive, but not reflexive. 
	Since all eight relations are defined in resolver logic,
	they can be decided using the model-checking algorithm for resolver logic.
	We also specialize our expressiveness and decidability results to specific value functions
	($\Sup$; $\Inf$; $\LimSup$; $\LimInf$) and to restricted classes of strategies (positional; finite-state).
	
	Third, we provide three more applications for resolver logic,
	in addition to checking strategic dominance.
	These applications show that resolvers play a central role in many different automata-theoretic problems.

	\smallskip\noindent
	{\color{lipicsGray}\bfseries Three more applications.~}
	Our first application concerns the {\em co-safety and co-liveness} of boolean and quantitative automata \cite{DBLP:conf/fossacs/HenzingerMS23,DBLP:conf/concur/BokerHMS23}.
	In~\cite{DBLP:conf/concur/BokerHMS23}, while the authors solved the problems of deciding safety or liveness, they left open the {\em bottom-value problem} for quantitative automata, which needs to be solved when checking if a quantitative automaton specifies a co-safety or co-liveness property.
	The bottom value of a quantitative automaton $\A$ can be defined as the infimum over all words $w$ of the
	supremum over all resolvers $f$ of the value~$\A^f(w)$.
	The bottom value, and all similarly defined values, can therefore be computed using the model-checking
	algorithm for resolver logic.
	
	Our second application concerns the {\em history-determinism} of boolean and quantitative automata.
	For a model $\B$ to strategically dominate~$\A$,
	different resolvers for $\A$ may give rise to different resolvers for~$\B$.
	Alternatively, one may postulate that in refinement,
	all resolvers for $\A$ should be ``dominated'' by the same resolver for~$\B$,
	that is, a single deterministic implementation of $\B$ should capture all possible deterministic
	implementations of~$\A$.
	Such {\em blind domination} gives rise to a nonreflexive relation on quantitative automata which implies
	simulation.
	We show that a quantitative automaton is history-deterministic iff the automaton blindly dominates itself.
	Consequently, our model-checking algorithm for resolver logic can be used to check history-determinism and
	generalizes the algorithm provided in \cite{DBLP:conf/csl/HenzingerP06} for checking the history
	determinism of boolean automata.
	
	Our third application concerns the {\em specification of hyperproperties}.
	A hyperproperty is a set of properties~\cite{DBLP:journals/jcs/ClarksonS10};
	hyperproperties occur in many different application contexts such as system security.
	A nondeterministic boolean automaton $\A$ can be viewed as specifying a hyperproperty $[\![A]\!]$ in the
	following natural way:
	if $R(\A)$ is the set of all possible resolvers for the nondeterminism of~$\A$,
	then $[\![\A]\!]=\{\A^f \st f\in R(\A)\}$,
	where $\A^f$ is the property (or ``language'') that is specified by $\A$ if its nondeterminism is resolved
	by~$f$.
	In the same way, nondeterministic quantitative automata can be used to specify quantitative hyperproperties as
	sets of quantitative languages~\cite{DBLP:journals/tocl/ChatterjeeDH10}.
	We show that there are simple automata that specify the boolean hyperproperties that
	contain all safety (resp.\ co-safety) properties,
	which cannot be specified in HyperLTL~\cite{DBLP:conf/post/ClarksonFKMRS14}.
	However, there are also boolean hyperproperties that can be specified in HyperLTL but not by
	applying resolvers to nondeterministic automata.
	Finally, we show how resolver logic can decide the inclusion problem for resolver-specified hyperproperties.

	\section{Definitional Framework}
	
	Let $\Sigma = \{a,b,\ldots\}$ be a finite alphabet of letters.
	An infinite (resp. finite) word (a.k.a. trace or execution) is an infinite (resp. finite) sequence of letters $w \in \Sigma^\omega$ (resp. $u \in \Sigma^*$).
	Given $u \in \Sigma^*$ and $w \in \Sigma^* \cup \Sigma^\omega$, we write $u \prec w$ when $u$ is a strict prefix of~$w$.
	We denote by $|w|$ the length of $w \in \Sigma^* \cup \Sigma^\omega$ and, given $a \in \Sigma$, by $|w|_a$ the number of occurrences of $a$ in $w$.
	We assume that the reader is familiar with formal language theory.	
	
	\begin{definition}[automaton]
		A \emph{(quantitative) automaton} is a tuple $\A = (\Sigma, Q, s, \Delta, \mu, \nu)$ where
		$\Sigma$ is a finite alphabet, $Q$ is a finite set of states, $s \in Q$ is the initial state, $\Delta \subseteq Q \times \Sigma \times Q$ is a transition relation, $\mu \colon \Delta \to \QQ$ is a weight function, and $\nu \colon \QQ^\omega \to \RR$ is a value function.
		The size of $\A$ is defined by $|Q|+\sum_{\delta\in\Delta} 1+\log_2(\mu(\delta))$.
		A \emph{run} $\pi$ over a finite (resp. an infinite) word $w = \sigma_0\sigma_1\dots$ is a finite (resp. an infinite) sequence $\pi = q_0 \sigma_0 q_1 \sigma_1 \dots$ such that $q_0 = s$ is the initial state of $\A$ and $(q_i, \sigma_i, q_{i+1})\in\Delta$ holds for all $0 \leq i < |w|$ (resp. $i\in\NN$).
		A \emph{history} is a run over a finite word, and we denote the set of histories of $\A$ by $\Pi_{\A}$.
		The weight sequence $\mu(\pi)$ of a run $\pi$ is defined by $x_0x_1\dots$ where $x_i = \mu(q_i, \sigma_i, q_{i+1})$ for all $i\in\NN$.
		The value of a run $\pi$ is defined as $\nu(\mu(\pi))$.
		In this paper, we consider automata with $\nu \in \{\Inf, \Sup, \LimInf, \LimSup\}$ and without loss of generality $\mu : \Delta \to \NN$ since a finite set of rational weights can be scaled and shifted to naturals and back.
	\end{definition}

	In general, automata resolve their nondeterminism by taking the $\sup$ over its set of runs on a given word.
	We take an alternative view and pair every automaton with a \emph{resolver}---an explicit description of how nondeterminism is resolved, which is a central concept in this work.
	Given a finite prefix of a run and the next input letter, a resolver determines the next state of the automaton.
	
	\begin{definition}[resolver]
		Let $\A = (\Sigma, Q, s, \Delta, \mu, \nu)$ be an automaton.
		A \emph{resolver} for $\A$ is a function $f \colon \Pi_{\A} \times \Sigma \to Q$ such that for every history $h = q_0 \sigma_0 q_1 \sigma_1 \ldots q_n \in \Pi_{\A}$ and every $\sigma \in \Sigma$ we have $(q_n, \sigma, f(h, \sigma)) \in \Delta$.
		A resolver $f$ for $\A$ and a word $w = \sigma_1 \sigma_2 \ldots \in \Sigma^\omega$ produce a unique infinite run $\pi_{f,w} = q_0 \sigma_1 q_1 \sigma_2 \ldots$ of $\A$ such that
		$q_0 = s$ and $f(q_0 \sigma_1 \ldots  q_{i-1}, \sigma_i) = q_i$ for all $i \geq 1$.
		Given an automaton $\A$, we denote by $R(\A)$ the set of its resolvers.
		Given a resolver $f\in R(\A)$, we define the quantitative language $\A^{f} \colon w \mapsto \nu(\mu(\pi_{f,w}))$ where $w \in \Sigma^\omega$.
		We also define the quantitative language
		$\A_{\sup} \colon w \mapsto \sup_{f\in R(\A)} \A^{f}(w)$, which is the standard interpretation of nondeterminism for automata.
	\end{definition}

	We define finite-memory and positional resolvers the usual way: a resolver $f$ is \emph{finite-memory} iff it can be implemented by a finite-state machine, and it is \emph{positional} iff the output of $f$ only depends on the last state in the input history and the incoming letter (see~\cite[Sec. 1.5]{DBLP:journals/corr/abs-2305-10546} for the formal definitions).
	Given an automaton $\A$, we denote by $R^{\text{fin}}(\A)$ the set of its finite-memory resolvers, and by $R^{\text{pos}}(\A)$ set of its positional resolvers.
	Let us demonstrate the notion of resolvers.
	
	\begin{figure}[t]\centering
		\scalebox{\picturescaletikz}{
		\begin{tikzpicture}[bg={\A}, node distance=.8cm]
			\node [state, initial, label=center:$q_0$] (q0) {};
			\node [right=of q0, state, label=center:$q_1$] (q1) {};
			\node [right=of q1, state, label=center:$q_2$] (q2) {};
			\path[transition]
			(q0) edge[loop above] node[above]{$\Sigma : 0$} (q0)
			(q0) edge node[above]{$a : 0$} (q1)
			(q1) edge[loop above] node[above]{$b : 1$} (q1)
			(q1) edge node[above]{$a : 0$} (q2)
			(q2) edge[loop above] node[above]{$\Sigma : 2$} (q2)
			;
		\end{tikzpicture}
		}
		\caption{A $\LimSup$-automaton $\A$ over $\Sigma = \{a,b\}$ with $R^{\text{pos}}(\A) \subsetneq R^{\text{fin}}(\A) \subsetneq R(\A)$.} \label{fig:resolvers}
	\end{figure}
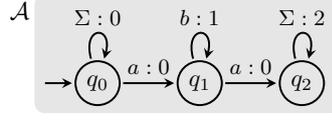
		
	\begin{example} \label{ex:resolvers}
		Let $\A$ be a $\LimSup$-automaton over the alphabet $\Sigma = \{a,b\}$ as in \cref{fig:resolvers} and observe that the only source of nondeterminism is the transitions $(q_0, a, q_0)$ and $(q_0, a, q_1)$.
		
		Consider a resolver $f_1$ that maps any history of the form $q_0 (b q_0)^*$ followed by $a$ to $q_1$.
		Intuitively, it is a \emph{positional} resolver because it ignores the transition $(q_0, a, q_0)$ and its output only depends on the current state and the next letter.
		It denotes the language $\A^{f_1}$ that maps a given word $w$ to 0 if $w$ has no $a$, to 1 if $w$ has exactly one $a$, and to 2 otherwise.
		
		Now, consider a resolver $f_2$ that maps the histories of the form $q_0 (b q_0)^*$ followed by $a$ to $q_0$, and those of the form $q_0 (b q_0)^* a q_0 (b q_0)^*$ followed by $a$ to $q_1$.
		Intuitively, it is a \emph{finite-state} resolver because it only distinguishes between $a$ occurring once or twice or neither.
		The language $\A^{f_2}$ then maps a word $w$ to 0 if $w$ has at most one $a$, to 1 if $w$ has exactly two $a$s, and to 2 otherwise.

		Finally, consider a resolver $f_3$ that maps every history that ends at $q_0$ with the incoming letter $a$ to $q_1$ if the given history is of the form $q_0 (\Sigma q_0)^* (b q_0)^p$ where $p$ is a prime, and to $q_0$ otherwise.
		Intuitively, it is an infinite-memory resolver because it needs to store the length of every block of $b$s, which is not bounded, and check whether it is prime.
		The language $\A^{f_3}$ maps a word $w$ to 0 if it has no prefix $u = v b^p a$ with $v \in \Sigma^*$, to 1 if it has such a prefix $u$ and $w = u b^\omega$, and to 2 otherwise.
	\end{example}

	Next, we introduce the notion of \emph{partial resolvers}.
	In contrast to (nonpartial) resolvers, partial resolvers output a set of successor states for a given history and letter.
	
	\begin{definition}[partial resolver]
		Let $\A = (\Sigma, Q, s, \Delta, \mu, \nu)$ be an automaton.
		A \emph{partial resolver} for $\A$ is a function $f\colon \Pi_{\A} \times \Sigma \to 2^{Q}$ such that for every history $h = q_0 \sigma_0 q_1 \sigma_1 \ldots q_n \in \Pi_{\A}$ and every $\sigma \in \Sigma$ we have $\{(q_n, \sigma, q) \st q \in f(h, \sigma) \} \subseteq \Delta$.
		A collection of partial resolvers $f_1, \dots, f_n$ for $\A$ is said to be \emph{conclusive} when $|\bigcap_{i=1}^n f_i(h, \sigma)| = 1$ for all  $h \in \Pi_{\A}$ and all $\sigma \in \Sigma$.
		Given a conclusive collection of resolvers $f_1, \dots, f_n$, we denote by $\A^{\{f_1, \dots, f_n\}}$ the quantitative language $\A^{f}$ where $f$ is a resolver defined by $f(h, \sigma) = \bigcap_{i=1}^n f_i(h, \sigma)$ for all  $h \in \Pi_{\A}$ and all $\sigma \in \Sigma$.
	\end{definition}

	Partial resolvers are particularly useful when we consider products of automata.
	In particular, we will use these objects to capture simulation-like relations in \cref{sec:resolverrelations}.

	\begin{definition}[synchronized product]
		Let $\A_1$ and $\A_2$ be two automata such that $\A_i = (\Sigma, Q_i, s_i, \Delta_i, \mu_i, \nu_i)$ for $i \in \{1,2\}$.
		For $k\in\{1, 2\}$, the \emph{(synchronized) product} $\A_1 \times_k \A_2$ corresponds to the input-synchronization of $\A_1$ and $\A_2$ where the transition weights are taken from $\A_k$.
		Formally, $\A_1 \times_k \A_2 = (\Sigma, Q_1 \times Q_2, (s_1, s_2), \Delta, \mu, \nu_k)$ where the transition relation is such that $((q_1, q_2), \sigma, (q'_1, q'_2)) \in \Delta$ if and only if $(q_i, \sigma, q'_i) \in\Delta_k$ for $i \in \{1,2\}$, and the weight function is such that $\mu((q_1, q_2), \sigma, (q'_1, q'_2)) = \mu_k(q_k, \sigma, q'_k)$.
		
		Given $i\in\{1, 2\}$, we denote by $R_i^S(\A_1, \A_2)$ the set of partial resolvers operating non-partially on the $i$th component of any product between $\A_1$ and $\A_2$.
		Formally, every $f \in R_i(\A_1, \A_2)$ is a partial resolver that satisfies the following:
		for all $h \in \Pi_{\A_1 \times_i \A_2}$, all $(q_1, q_2) \in Q_1 \times Q_2$ and all $\sigma \in \Sigma$, there exists $(q_i, \sigma, q'_i) \in \Delta_i$ such that $f(h \cdot (q_1, q_2), \sigma)) = \{(q'_1, q'_2) \st (q_{3-i}, \sigma, q'_{3-i}) \in \Delta_{3-i}\}$.		
	\end{definition}
	
	Let us demonstrate the notions of partial resolvers and synchronized products.
	
	\begin{figure}[t]\centering
		\noindent\begin{minipage}{.24\linewidth}\centering
			\scalebox{\picturescaletikz}{
			\begin{tikzpicture}[bg={\A}, node distance=.8cm]
				\node [state, initial, label=center:$s_0$] (q0) {};
				\node [right=of q0, state, label=center:$s_1$] (q1) {};
				\path[transition]
				(q0) edge[loop above] node[above]{$\Sigma : 0$} (q0)
				(q0) edge node[above]{$a : 0$} (q1)
				(q1) edge[loop above] node[above]{$\Sigma : 1$} (q1)
				;
			\end{tikzpicture}
			}
		\end{minipage}
		\noindent\begin{minipage}{.24\linewidth}\centering
			\scalebox{\picturescaletikz}{
			\begin{tikzpicture}[bg={\B}, node distance=.8cm]
				\node [state, initial, label=center:$q_0$] (q0) {};
				\node [right=of q0, state, label=center:$q_1$] (q1) {};
				\path[transition]
				(q0) edge[loop above] node[above]{$\Sigma : 1$} (q0)
				(q0) edge node[above]{$a : 0$} (q1)
				(q1) edge[loop above] node[above]{$\Sigma : 0$} (q1)
				;
			\end{tikzpicture}
			}
		\end{minipage}
		\noindent\begin{minipage}{.50\linewidth}\centering
			\scalebox{\picturescaletikz}{
			\begin{tikzpicture}[bg={$\A  \times_1 \B$}, node distance=.8cm]
				\tikzstyle{statepair}=[thick,minimum height=18pt, rectangle, rounded corners, draw]
				\node [statepair] (q0) {$s_0,q_0$};
				\draw[transition](q0.205)++(205:13pt) -- (q0);
				\node [statepair, right=of q0] (q1) {$s_1,q_0$};
				\node [statepair, below=of q0] (q2) {$s_0,q_1$};
				\node [statepair, below=of q1] (q3) {$s_1,q_1$};
				\path[transition]
				(q0) edge[out=190,in=170, min distance=4mm, looseness=7] node[left]{$\Sigma : 0$} (q0)
				(q1) edge[out=10,in=350, min distance=4mm, looseness=7] node[right]{$\Sigma : 1$} (q1)
				(q2) edge[out=190,in=170, min distance=4mm, looseness=7] node[left]{$\Sigma : 0$} (q2)
				(q3) edge[out=10,in=350, min distance=4mm, looseness=7] node[right]{$\Sigma : 1$} (q3)
				(q0) edge node[above]{$a : 0$} (q1)
				(q0) edge node[left]{$a : 0$} (q2)
				(q0) edge node[sloped, above]{$a : 0$} (q3)
				(q1) edge node[right]{$a : 1$} (q3)
				(q2) edge node[above]{$a : 0$} (q3)
				;
			\end{tikzpicture}
			}
		\end{minipage}
	\caption{Two $\LimSup$-automata $\A$ and $\B$ and the product $\A \times_1 \B$.}\label{fig:product}
	\end{figure}
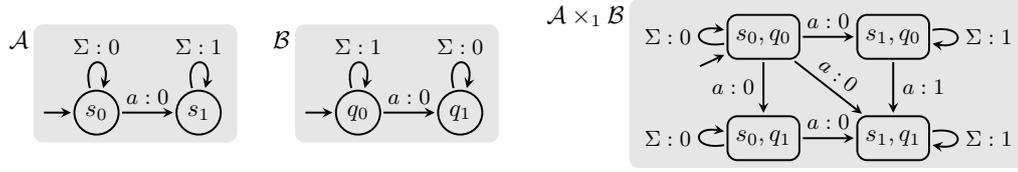
	
	\begin{example} \label{ex:product}
		Let $\A$ and $\B$ be as in \cref{fig:product}.
		Because of the loop in their initial states, as in \cref{ex:resolvers}, both automata have infinitely many resolvers, 
		Their product $\A \times_1 \B$ is also shown in \cref{fig:product}.
		Intuitively, the product is similar to the boolean construction with difference of handling the transition weights instead of the accepting states.
		For $\A \times_1 \B$, the transition weights are obtained from the corresponding ones in $\A$.
		
		Consider the positional resolver $f$ of $\A$ that moves from $s_0$ to $s_1$ as soon as $a$ occurs.
		Observe that there is a partial resolver $f'$ of $\A \times_1 \B$, namely $f' \in R_1(\A, \B)$, that imitates $f$.
		In particular, $f'((s_0, q_0), a) = \{(s_1, q_0), (s_1, q_1)\}$ since $f$ moves $\A$ from $s_0$ to $s_1$ with $a$, but it is not specified how the $\B$-component of the product resolves this nondeterministic transition.
		Now, suppose we have $g' \in R_2(\A, \B)$ such that $g'((s_0, q_0), a) = \{(s_0, q_1), (s_1, q_1)\}$.
		The two partial resolvers $f'$ and $g'$ together are conclusive for $\A \times_1 \B$, as witnessed by $f'((s_0, q_0), a) \cap g'((s_0, q_0), a) = \{(s_1,q_1)\}$.
	\end{example}

	\begin{remark} \label{rm:productresolvers}
		As demonstrated in \cref{ex:product}, when we consider partial resolvers over product automata, the partial resolvers operating on the distinct components are conclusive for the product.
		Formally, let $\A$ and $\B$ be two automata and consider one of their products.
		Every pair of partial resolvers $f \in R_1(\A, \B)$ and $g \in R_2(\A, \B)$ is conclusive for the product automaton by definition, and thus corresponds to a (non-partial) resolver over the product.
	\end{remark}

	\section{Strategic Dominance and Other Resolver-Based Relations} \label{sec:resolverrelations}
	
\begin{figure}\centering
		\noindent\begin{minipage}{.75\linewidth}\centering
			\scalebox{.9}{
			$$\fbox{$\begin{array}{ll}
					\textbf{Notation} & \textbf{Problem statement}
					\smallskip\\
					\A \inclusion \B & \forall w \in \Sigma^\omega: \A_{\sup}(w) \leq \B_{\sup}(w)
					\smallskip\\
					\A \simulation \B & 
					\begin{array}{l}
						\text{no winning strategy for Challenger} \\ \text{\qquad in the simulation game for $\A$ and $\B$}
					\end{array}
					\smallskip\\
					\A \commitsim\B &\forall f \in R(\A): \exists g \in R(\B)  :\forall w \in \Sigma^\omega : \A^{f}(w) \leq \B^{g}(w)
					\smallskip\\
					\A \blindsim \B & \exists g \in R(\B): \forall f \in R(\A): \forall w \in \Sigma^\omega: \A^{f}(w) \leq \B^{g}(w)
					\smallskip\\
					\A \weakcommitsim \B & \forall w \in \Sigma^\omega:  \forall f \in R(\A): \exists g \in R(\B):\A^{f}(w) \leq \B^{g}(w)
					\smallskip\\
					\A\weakblindsim  \B & \forall w \in \Sigma^\omega: \exists g \in R(\B): \forall f \in R(\A): \A^{f}(w) \leq \B^{g}(w)
					\smallskip\\
					\A\commitsim_{\times}\B & 
					\begin{array}{l}
						\forall f\in R_1(\A, \B):\exists g\in R_2(\A, \B):\forall w\in\Sigma^\omega\\
						\qquad (\A\times_1\B)^{\{f, g\}}(w) \leq (\A\times_2\B)^{\{f, g\}}(w)
					\end{array}
					\smallskip\\  
					\A\blindsim_{\times}\B & 
					\begin{array}{l}
						\exists g\in R_2(\A, \B):\forall f\in R_1(\A, \B):\forall w\in\Sigma^\omega\\
						\qquad (\A\times_1\B)^{\{f, g\}}(w) \leq (\A\times_2\B)^{\{f, g\}}(w)
					\end{array}
					\smallskip\\ 
					\A\weakcommitsim_{\times}\B & 
					\begin{array}{l}
						\forall w\in\Sigma^\omega:\forall f\in R_1(\A, \B):\exists g\in R_2(\A, \B)\\
						\qquad (\A\times_1\B)^{\{f, g\}}(w) \leq (\A\times_2\B)^{\{f, g\}}(w)
					\end{array}
					\smallskip\\
					\A\weakblindsim_{\times}\B & 
					\begin{array}{l}
						\forall w\in\Sigma^\omega:\exists g\in R_2(\A, \B):\forall f\in R_1(\A, \B)\\
						\qquad (\A\times_1\B)^{\{f, g\}}(w) \leq (\A\times_2\B)^{\{f, g\}}(w)
					\end{array}
					\smallskip\\
				\end{array}$}$$
			}
		\end{minipage}
		\noindent\begin{minipage}{.24\linewidth}\centering
			\scalebox{\picturescaletikz}{
			\begin{tikzpicture}[bg={}, node distance=0.8cm]
				\tikzstyle{class}=[thick,minimum width=36pt, rectangle, rounded corners, draw]
				\node[class] (inclusion) {$\weakblindsim, \weakcommitsim, \weakblindsim_{\times}, \weakcommitsim_{\times}, \inclusion$};
				\node[class, below =of inclusion] (commitsim) {$\commitsim$};
				\node[class, below =of commitsim] (commitsimprod) {$\commitsim_{\times}, \simulation$};
				\node[class, below =of commitsimprod] (blindsimprod) {$\blindsim_{\times}$};
				\node[class, below =of blindsimprod] (blindsim) {$\blindsim$};
				
				\path[transition]
				(blindsim) edge (blindsimprod)
				(blindsimprod) edge (commitsimprod)
				(commitsimprod) edge (commitsim)
				(commitsim) edge (inclusion)
				;
			\end{tikzpicture}
			}
	\end{minipage}
		\caption{Left: The definitions of inclusion (denoted ${\inclusion}$), simulation (denoted ${\simulation}$), and the resolver relations we study in \cref{sec:resolverrelations} including strategic dominance (denoted ${\commitsim}$). Right: The implications between these relations as proved in \cref{cl:inclusion,cl:simulation,cl:implications,cl:restrictedresolvers,cl:counterexamples,cl:corollary}.} 
		\label{fig:relations}
	\end{figure}
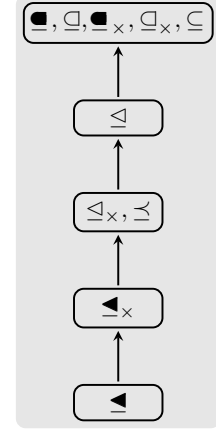

	Given two automata $\A$ and $\B$, we investigate the relations between the problems defined in \cref{fig:relations}.
	We denote the \emph{strategic dominance} relation by ${\commitsim}$: the automaton $\A$ is strategically dominated by $\B$, denoted $\A \commitsim \B$, iff for all resolvers $f \in R(\A)$ there exists a resolver $g \in R(\B)$ such that $\A^f(w) \leq \B^g(w)$ for all words $w \in \Sigma^\omega$.
	Intuitively, this holds when each deterministic implementation of $\A$ can be countered by some deterministic implementation of $\B$ that provides a value at least as large for each word.	
	
	We define and study other resolver-based relations and compare their expressiveness.
	In ${\weakcommitsim}$, the automaton $\B$ has the freedom to choose a resolver per input word (unlike in ${\commitsim}$).
	The relations ${\blindsim}$ and ${\weakblindsim}$ can be seen as variants of ${\commitsim}$ and ${\weakcommitsim}$ where the resolvers of $\B$ are \emph{blind}, meaning that they cannot depend on the resolvers of $\A$.
	
	To capture simulation-like relations, we additionally define the relations ${\commitsim_{\times}}, {\blindsim_{\times}}, {\weakcommitsim_{\times}}, {\weakblindsim_{\times}}$ that relate product automata through their partial resolvers.
	In particular, we show that ${\commitsim_{\times}}$, the product-based strategic dominance, coincides with simulation.
	Thanks to \cref{rm:productresolvers}, we are able to reason about these relations the same way we do for non-partial resolvers.

	In addition to these resolver relations, we denote the trace-inclusion preorder by ${\inclusion}$ and the simulation by ${\simulation}$.
	To define simulation formally, let us recall quantitative simulation games~\cite{DBLP:journals/tocl/ChatterjeeDH10}:
	Let $\A = (\Sigma, Q_{\A}, s_{\A}, \Delta_{\A}, \mu_{\A}, \nu_{\A})$ and $\B = (\Sigma, Q_{\B}, s_{\B}, \Delta_{\B}, \mu_{\B}, \nu_{\B})$.
	A strategy $\tau$ for Challenger is a function from $(Q_{\A} \times Q_{\B})^*$ to $\Sigma \times Q_{\A}$ satisfying for all $\pi = (q_1, p_1) \ldots (q_n, p_n) \in (Q_{\A} \times Q_{\B})^*$, if $\tau(\pi) = (\sigma, q)$ then $(q_n, \sigma, q) \in \Delta_{\A}$.
	Given a strategy $\tau$ for Challenger, the set of outcomes is the set of pairs $(q_0 \sigma_1 q_1 \sigma_2 q_2 \ldots, p_0 \sigma_1 p_1 \sigma_2 p_2 \ldots)$ of runs such that $q_0 = s_{\A}$, $p_0 = s_{\B}$, and for all $i \geq 0$ we have $(\sigma_{i+1}, q_{i+1}) = \tau((q_0, p_0) \ldots (q_i, p_i))$ and $(p_i, \sigma_i, p_{i+1}) \in \Delta_{\B}$.
	A strategy $\tau$ for Challenger is winning iff $\nu_{\A}(\mu_{\A}(r_1)) > \nu_{\B}(\mu_{\B}(r_2))$ for all outcomes $(r_1, r_2)$ of $\tau$.
	
	Given a problem instance $\A \sim \B$ where ${\sim}$ is one of the relations in \cref{fig:relations}, we write $\A \sim^{\text{fin}} \B$ (resp. $\A \sim^{\text{pos}} \B$) for the restriction of the corresponding problem statement to finite-memory (resp. positional) resolvers.
	For example, $\A \commitsim^{\text{fin}} \B$ iff for all $f \in R^{\text{fin}}(\A)$ there exists $g \in R^{\text{fin}}(\B)$ such that $\A^{f}(w) \leq \B^{g}(w)$ for all $w \in \Sigma^\omega$.

	\begin{remark}
		The results in this paper hold for quantitative $\Inf$-, $\Sup$-, $\LimSup$-, and $\LimInf$-automata as well as boolean safety, reachability, Büchi, and coBüchi automata.
		Moreover, they also hold when restricted to finite-state or positional resolvers.
	\end{remark}
	
	We start with a short lemma showing that the supremum over the values of any word $w$ is attainable by some run over $w$, which follows from the proof of \cite[Thm.~3]{DBLP:journals/tocl/ChatterjeeDH10}.
	
	\begin{lemma}\label{cl:supsupsup}
	Let $\nu \in \{\Inf, \Sup, \LimSup, \LimInf\}$ and $\A$ be a $\nu$-automaton.
	For every $w \in \Sigma^\omega$ there exist $f \in R(\A)$ such that $\A^{f}(w) = \A_{\sup}(w)$.
	\end{lemma}

	\subsection{Implications Between Resolver Relations} \label{sec:implications}
	
	We now prove the implications in \cref{fig:relations} without any memory constraints on the resolvers.
	We start with showing that some of these relations coincide with inclusion.
	
	\begin{proposition} \label{cl:inclusion}
	Let $\nu_1, \nu_2 \in \{\Inf, \Sup, \LimSup, \LimInf\}$.
	For all $\nu_1$-automata $\A$ and all $\nu_2$-automata $\B$, we have $\A \weakcommitsim \B$ iff $\A \weakblindsim \B$ iff $\A \weakcommitsim_{\times} \B$ iff $\A \weakblindsim_{\times} \B$ iff $\A \inclusion \B$.
	This is equally true for boolean safety, reachability, Büchi, and coBüchi automata.
	\end{proposition}

	Next, we show an equivalent formulation for simulation.
	
	\begin{proposition} \label{cl:simulation}
	Let $\nu_1, \nu_2 \in \{\Inf, \Sup, \LimSup, \LimInf\}$.
	For all $\nu_1$-automata $\A$ and all $\nu_2$-automata $\B$, we have $\A \commitsim_{\times} \B$ iff $\A \simulation \B$.
	This is equally true for boolean safety, reachability, Büchi, and coBüchi automata.
	\end{proposition}

	We proceed to show the implications posed in \cref{fig:relations}.
	
	\begin{proposition} \label{cl:implications}
	Let $\nu_1, \nu_2 \in \{\Inf, \Sup, \LimSup, \LimInf\}$.
	For all $\nu_1$-automata $\A$ and all $\nu_2$-automata $\B$, the following statements hold.
	Moreover, they are equally true for boolean safety, reachability, Büchi, and coBüchi automata.
	\begin{multicols}{2}
		\begin{enumerate}
			\item $\A \commitsim \B \Rightarrow \A \inclusion \B$
			\item $\A \simulation \B \Rightarrow \A \commitsim \B$
			\item $\A \blindsim_{\times} \B \Rightarrow \A \simulation \B$
			\item $\A \blindsim \B \Rightarrow \A \blindsim_{\times} \B$
		\end{enumerate}
	\end{multicols}
	\end{proposition}
	
	The implications given in \cref{fig:relations} (and proved in \cref{cl:inclusion,cl:simulation,cl:implications})
	also hold when the problem statements are restricted to only finite-memory resolvers or positional resolvers.
	
	\begin{proposition} \label{cl:restrictedresolvers}
	Let $\nu_1, \nu_2 \in \{\Inf, \Sup, \LimSup, \LimInf\}$.
	For all $\nu_1$-automata $\A$ and all $\nu_2$-automata $\B$ and each $r \in \{\text{fin}, \text{pos}\}$ the following statements hold.
	Moreover, they are equally true for boolean safety, reachability, Büchi, and coBüchi automata.
	\begin{enumerate}
		\item $\A \weakcommitsim^{r} \B \Leftrightarrow \A \weakblindsim^{r} \B \Leftrightarrow \A \weakcommitsim_{\times}^{r} \B \Leftrightarrow \A \weakblindsim_{\times}^{r} \B \Leftrightarrow \A \inclusion^{r} \B$
	\end{enumerate}
	\vspace{-0.65cm}
	\begin{multicols}{2}
	\begin{enumerate}[resume]
		\item $\A \commitsim_{\times}^{r} \B \Leftrightarrow \A \simulation^{r} \B$
		\item $\A \commitsim^{r} \B \Rightarrow \A \inclusion^{r} \B$
		\item $\A \simulation^{r} \B \Rightarrow \A \commitsim^{r} \B$
		\item $\A \blindsim_{\times}^{r} \B \Rightarrow \A \simulation^{r} \B$
		\item $\A \blindsim^{r} \B \Rightarrow \A \blindsim_{\times}^{r} \B$
		\item[] 
	\end{enumerate}
	\end{multicols}
	\end{proposition}
	
	\subsection{Separating Examples for Resolver Relations} \label{sec:counterexamples}
	
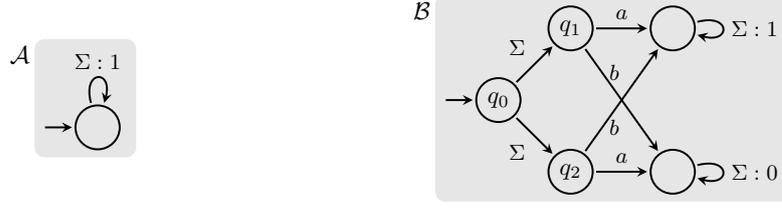
\begin{figure}[t]\centering
		\noindent\begin{minipage}{.49\linewidth}\centering
			\scalebox{\picturescaletikz}{
			\begin{tikzpicture}[bg={\A}, node distance=.8cm]
				\node [state, initial] (q0) {};
				\path[transition]
				(q0) edge[loop above] node {$\Sigma : 1$} (q0)
				;
			\end{tikzpicture}
			}
		\end{minipage}
		\noindent\begin{minipage}{.49\linewidth}\centering
			\scalebox{\picturescaletikz}{
			\begin{tikzpicture}[bg={\B}, node distance=.8cm]
				\node [state, initial, label=center:$q_0$] (q0) {};
				\node [above right=of q0, state, label=center:$q_1$] (q1) {};
				\node [below right=of q0, state, label=center:$q_2$] (q2) {};
				\node [right=of q1, state] (q3) {};
				\node [right=of q2, state] (q4) {};
				\path[transition]
				(q0) edge node[above left]{$\Sigma$} (q1)
				(q0) edge node[below left]{$\Sigma$} (q2)
				(q1) edge node[above]{$a$} (q3)
				(q1) edge[pos=.4] node[above]{$b$} (q4)
				(q2) edge[pos=.4] node[below]{$b$} (q3)
				(q2) edge node[above]{$a$} (q4)
				(q3) edge[loop right] node[right]{$\Sigma : 1$} (q3)
				(q4) edge[loop right] node[right]{$\Sigma : 0$} (q4)
				;
			\end{tikzpicture}
			}
		\end{minipage}
	\caption{Two automata $\A$ and $\B$ such that $\A \inclusion^{\text{pos}} \B$ but $\A \not\commitsim^{\text{pos}} \B$. Note that $\A$ and $\B$ can be $\Inf$-, $\Sup$-, $\LimSup$-, or $\LimInf$-automata as well as safety, reachability, Büchi, or coBüchi automata. The exact values of the omitted weights depend on the considered value function. For example, we can take as weight 1 for $\Inf$ and 0 for $\Sup$ while both choices work for $\LimSup$ and $\LimInf$.}\label{fig:sep1}
	\end{figure}

	In this part, we provide separating examples for the implications we proved above, establishing a hierarchy of relations given in \cref{fig:relations}.
	The counter-examples we used in the proofs below are displayed in \cref{fig:sep1,fig:sep2sep4}.
	
	\begin{proposition} \label{cl:counterexamples}
	Let $\nu_1, \nu_2 \in \{\Inf, \Sup, \LimSup, \LimInf\}$.
	For each statement below, there exist a $\nu_1$-automaton $\A$ and a $\nu_2$-automaton $\B$ to satisfy it.
	Moreover, they are equally true for boolean safety, reachability, Büchi, and coBüchi automata.
	\begin{multicols}{2}
		\begin{enumerate}
			\item $\A \inclusion^{\text{pos}} \B \land \A \not\commitsim^{\text{pos}} \B$
			\item $\A \commitsim^{\text{pos}} \B \land \A \not\simulation^{\text{pos}} \B$
			\item $\A \simulation^{\text{pos}} \B \land \A \not\blindsim_{\times}^{\text{pos}} \B$
			\item $\A \blindsim_{\times}^{\text{pos}} \B \land \A \not\blindsim^{\text{pos}} \B$
		\end{enumerate}
	\end{multicols}
	\end{proposition}

	\begin{figure}[t]\centering
		\noindent\begin{minipage}{.49\linewidth}\centering
			\scalebox{\picturescaletikz}{
			\begin{tikzpicture}[bg={\A}, node distance=.8cm]
				\node [state, initial, label=center:$s_0$] (q0) {};
				\node [right=of q0, state, label=center:$s_1$] (q1) {};
				\node [above right=of q1, state, label=center:$s_2$] (q2) {};
				\node [below right=of q1, state, label=center:$s_3$] (q3) {};
				\node [right=of q2, state] (q4) {};
				\node [right=of q3, state] (q5) {};
				\path[transition]
				(q0) edge node[above]{$a$} (q1)
				(q1) edge node[above left]{$a$} (q2)
				(q1) edge node[below left]{$a$} (q3)
				(q2) edge node[above]{$a$} (q4)
				(q2) edge[pos=.4] node[above]{$b$} (q5)
				(q3) edge[pos=.4] node[below]{$b$} (q4)
				(q3) edge node[above]{$a$} (q5)
				(q4) edge[loop right] node[right]{$\Sigma : 1$} (q4)
				(q5) edge[loop right] node[right]{$\Sigma : 0$} (q5)
				;
			\end{tikzpicture}
			}
		\end{minipage}
		\noindent\begin{minipage}{.49\linewidth}\centering
			\scalebox{\picturescaletikz}{
			\begin{tikzpicture}[bg={\B}, node distance=.8cm]
				\node [state, initial, label=center:$q_0$] (q0) {};
				\node [above right=of q0, state, label=center:$q_1$] (q1) {};
				\node [below right=of q0, state, label=center:$q_2$] (q2) {};
				\node [right=of q1, state, label=center:$q_3$] (q3) {};
				\node [right=of q2, state, label=center:$q_4$] (q4) {};
				\node [right=of q3, state] (q5) {};
				\node [right=of q4, state] (q6) {};
				\path[transition]
				
				(q0) edge node[above left]{$a$} (q1)
				(q0) edge node[below left]{$a$} (q2)
				(q1) edge node[above]{$a$} (q3)
				(q2) edge node[above]{$a$} (q4)
				(q3) edge node[above]{$a$} (q5)
				(q3) edge[pos=.4] node[above]{$b$} (q6)
				(q4) edge[pos=.4] node[below]{$b$} (q5)
				(q4) edge node[above]{$a$} (q6)
				(q5) edge[loop right] node[right]{$\Sigma : 1$} (q5)
				(q6) edge[loop right] node[right]{$\Sigma : 0$} (q6)
				;
			\end{tikzpicture}
			}
		\end{minipage}
	\caption{Two automata $\A$ and $\B$ such that 
		(i) $\A \commitsim^{\text{pos}} \B$ but $\A \not\simulation^{\text{pos}} \B$,
		(ii) $\A \simulation^{\text{pos}} \A$ but $\A \not\blindsim_{\times}^{\text{pos}} \A$, and
		(iii) $\B \blindsim_{\times}^{\text{pos}} \A$ but $\B \not\blindsim^{\text{pos}} \A$.
		The transitions that are not shown lead to a sink state with a self-loop on every letter with weight 0.
		Note that $\A$ and $\B$ can be $\Inf$-, $\Sup$-, $\LimSup$-, or $\LimInf$-automata as well as safety, reachability, Büchi, or coBüchi automata. The exact values of the omitted weights depend on the considered value function. For example, we can take as weight 1 for $\Inf$ and 0 for $\Sup$ while both choices work for $\LimSup$ and $\LimInf$.}\label{fig:sep2sep4}
	\end{figure}
	
	Since $R^\text{pos}(\A) = R^\text{fin}(\A) = R(\A)$ for each automaton $\A$ we consider in this section (see \cref{fig:sep1,fig:sep2sep4}), the statements above also hold for the finite-memory and the general cases.
	
	\begin{corollary} \label{cl:corollary}
	Let $\nu_1, \nu_2 \in \{\Inf, \Sup, \LimSup, \LimInf\}$.
	For each statement below, there exist a $\nu_1$-automaton $\A$ and a $\nu_2$-automaton $\B$ to satisfy it.
	Moreover, they are equally true for boolean safety, reachability, Büchi, and coBüchi automata.
	\begin{multicols}{2}
		\begin{enumerate}
			\item $\A \inclusion^{\text{fin}} \B \land \A \not\commitsim^{\text{fin}} \B$
			\item $\A \commitsim^{\text{fin}} \B \land \A \not\simulation^{\text{fin}} \B$
			\item $\A \simulation^{\text{fin}} \B \land \A \not\blindsim_{\times}^{\text{fin}}\B$
			\item $\A \blindsim_{\times}^{\text{fin}} \B \land \A \not\blindsim^{\text{fin}} \B$
			
			\item $\A \inclusion \B \land \A \not\commitsim \B$
			\item $\A \commitsim \B \land \A \not\simulation \B$
			\item $\A \simulation \B \land \A \not\blindsim_{\times} \B$
			\item $\A \blindsim_{\times} \B \land \A \not\blindsim \B$
		\end{enumerate}
	\end{multicols}
	\end{corollary}


	\begin{remark}
		The relations ${\blindsim}$ and ${\blindsim_{\times}}$ are not reflexive (and thus not a preorder).
		For the automaton $\A$ given in \cref{fig:sep2sep4}, we have $\A \not\blindsim_{\times}^{\text{pos}} \A$ as shown in the proof of Item (3) of \cref{cl:counterexamples}, and thus $\A \not\blindsim_{\times} \A$ since all of its resolvers are positional.
		Then, by Item (4) of \cref{cl:implications}, we also have $\A \not\blindsim \A$.
	\end{remark}

	\section{Resolver Logic} \label{sec:algorithm}
	We dedicate this section to describing resolver logic, which intuitively extends Presburger arithmetic by introducing variables evaluated by automata parameterized by quantified words and resolvers.
	We formally define resolver logic and show that the model-checking of a resolver logic formula is decidable.
	
	\begin{definition}[resolver logic]
		Let $\textsf{QA} = \{\A_1, \ldots, \A_n\}$ be a finite set of automata over the same alphabet $\Sigma$.
		For all $k\in\{1, \dots, n\}$, let $F_{k}$ be a set of resolver variables ranging over $R(\A_k)$, and let $W \subseteq \Sigma^\omega$ be a set of word variables ranging over $\Sigma^\omega$.
		We define the set $V=\{v_{(x,y)} \st x\in W, y \in \bigcup_{k=1}^n F_k\}$ of variables ranging over non-negative integers.
		A \emph{resolver logic formula} on the automata domain $\textsf{QA}$ is a term generated by the grammar $\varPsi ::= \exists x : \varPsi \mid \forall x : \varPsi \mid \varphi$, where $x\in W \cup \bigcup_{k=1}^n F_k$ and $\varphi \in \exists\textsf{FO}(\NN, =, +, 1)$ is an existential Presburger formula whose set of free variables is $V$.
	\end{definition}
%
	
	We write $|\varPsi|$ to denote the size of $\varPsi$ defined as $|\varPsi| = n + m + |\varphi|$ where $n = |\textsf{QA}|$ is the cardinality of the automata domain, $m=|W|+\sum_{i=1}^n |F_i|$ is the number of word and resolver variables in $\varPsi$, and $|\varphi|$ is the number of (existential) quantifiers in $\varphi$.
	Note that the strategic dominance and the other relations defined in \cref{sec:resolverrelations} are examples of resolver logic formulas.
	An assignment $\alpha$ maps variables of $W$ to words in $\Sigma^\omega$, variables of $F_k$ to resolvers in $R(\A_k)$ for all $k\in\{1, \dots, n\}$, and variables of $v \in V$ to values in $\NN$.
	In particular, $\alpha(v_{(x, x')}) = \A_k^f(w)$ where $x \in W$ and $x' \in F_k$ such that $\alpha(x) = w\in \Sigma^\omega$ and $\alpha(x') = f\in R(\A_k)$.
	The semantics of $\varPsi$ is defined as follows.
	\begin{align*} 
		&(\textsf{QA}, \alpha)  \models \varphi  \text{ iff $\varphi[\forall v \in V : v \gets \alpha(v)]$ holds} \\
		&(\textsf{QA}, \alpha)  \models \exists x \in W : \Psi  \text{ iff for some $w \in \Sigma^\omega$ we have $\alpha[x \gets w] \models \Psi$} \\
		&(\textsf{QA}, \alpha)  \models \forall x \in W : \Psi  \text{ iff for all $w \in \Sigma^\omega$ we have $\alpha[x \gets w] \models \Psi$} \\
		&(\textsf{QA}, \alpha)  \models \exists y \in F_k : \Psi  \text{ iff for some $f \in R(\A_k)$ we have $\alpha[y \gets f] \models \Psi$}\\
		&(\textsf{QA}, \alpha)  \models \forall y \in F_k : \Psi  \text{ iff for all $f \in R(\A_k)$ we have $\alpha[y \gets f] \models \Psi$}
	\end{align*}



	\begin{theorem} \label{cl:algo}
		The model-checking of a given resolver logic formula $\varPsi$ is decidable.
		When $|\varPsi|$ is fixed, model-checking is in $d$-$\ExpTime$ if there are $d>0$ quantifier alternations and $\PTime$ if there is no quantifier alternation.
	\end{theorem}

	The decision procedure relies on the following:
	(1) Each variable assignment can be represented by a single tree.
	(2) Each resolver logic formula admits a parity tree automata that accepts all tree-encoded assignments that satisfy its inner Presburger formula.
	(3) The model-checking of a given resolver formula can be decided based on nested complementations and projections over the above parity tree automaton.
	Below we provide an overview of the corresponding constructions (the full proof is deferred to the appendix due to space constraints).
	We start by defining trees and describing to use them to encode assignments.

	\smallskip\noindent
	{\color{lipicsGray}\bfseries Trees.~}
	Let $\Sigma$ be an alphabet for the structure of trees.
	We view the set $\Sigma^*$ of finite words as the domain of an infinite $|\Sigma|$-ary tree.
	The root is the empty word $\varepsilon$, and for a node $u \in \Sigma^*$ together with some letter $\sigma \in \Sigma$ we call $u \sigma$ the $\sigma$-successor of $u$.
	Let $\Lambda$ be an alphabet for the labeling of nodes.
	An infinite $\Lambda$-labeled $\Sigma$-structured tree is a function $t \colon \Sigma^* \to \Lambda$.
	We denote by $\Lambda^\omega_{\Sigma}$ the set of all such trees.
	In a tree $t\in \Lambda^\omega_{\Sigma}$, the word $w=\sigma_{0}\sigma_{1}\dots \in\Sigma^\omega$ induces the branch $t(w)$ defined as the infinite sequence $t(\varepsilon)\sigma_{0}t(\sigma_{0})\sigma_{1}t(\sigma_{0}\sigma_{1})\dots \in (\Lambda \times \Sigma)^\omega$.

	Let $\textsf{QA} = \{\A_1, \ldots, \A_n\}$ be a set of automata over the same alphabet $\Sigma$.
	Consider the resolver logic formula over $\textsf{QA}$ of the form $\varPsi = \nabla_1 x_1 : \dots : \nabla_m x_m : \varphi$, where $\nabla_i \in \{\exists, \forall\}$.
	The decision procedure encodes assignments for resolver and word variables of $\varPsi$ into $\Lambda$-labeled $\Sigma$-structured trees, where $\Lambda = \big(\{0,1\} \cup \bigcup_{i=1}^n Q_n\big)^m$.
	All dimensions of the tree that correspond to a word must have exactly one branch labeled by 1 (which encodes the word), and all other nodes are labeled by 0.
	Formally, the assignment $\alpha_t$ encoded by a tree $t\in \Lambda^\omega_{\Sigma}$ maps the word variable $x_j\in W$ to the unique word $\alpha_t(x_j)=\sigma_1\sigma_2\dots \in\Sigma^\omega$ for which the $j$th dimension of the branch $t(\alpha_t(x_j))$ is $t_j(\alpha_t(x_j))= 1 \sigma_1 1 \sigma_2\dots \in (\{1\} \times \Sigma)^\omega$.
	All dimensions of the tree that correspond to a resolver of $\A_k$ must respect its transition relation.
	Formally, the assignment $\alpha_t$ encoded by a tree $t\in \Lambda^\omega_{\Sigma}$ maps the resolver variable $x_i\in F_k$ to the unique resolver $\alpha_t(x_i)\in R(\A_k)$ defined by $\alpha_t(x_i)(\pi_k, \sigma) = t_i(u\sigma)$ where $\pi_k$ is the finite run of $\A_k^{\alpha_t(x_i)}$ over $u\in \Sigma^*$.
	Consequently, for all $x_i\in F_k$ and all $x_j\in W$, 
	the assignment $\alpha_t$ encoded by a tree $t\in \Lambda^\omega_{\Sigma}$ maps the variable $v_{(x_i,x_j)}\in V$ to the unique non-negative integer $\A_k^{\alpha_t(x_i)}(\alpha_t(x_j))$.
	Next, we describe the parity tree automaton constructed in the decision procedure.

	\smallskip\noindent
	{\color{lipicsGray}\bfseries Parity Tree Automata.~}
	A (nondeterministic) parity tree automaton $\T$ over $\Lambda^\omega_{\Sigma}$ is a tuple $(\Lambda, \Sigma, Q, I, \Delta, \theta)$ where $\Lambda$ is a finite labeling alphabet, $\Sigma$ is a finite structure alphabet, $Q$ is a finite set of states, $I \subseteq Q$ is a set of initial states, $\Delta \subseteq Q \times \Lambda \times (\Sigma \to Q)$ is a transition relation, and $\theta\colon Q \to \NN$ is the priority function.
	Note that the arity of the trees is $|\Sigma|$ and is statically encoded in the transition relation.
	A run of $\T$ over $t \in \Lambda^\omega_{\Sigma}$ is a $Q$-labeled $\Sigma$-structured tree $\pi \in Q^\omega_{\Sigma}$ 
	such that $\pi(\varepsilon)\in I$ 
	and for each $u\in\Sigma^*$ we have $(\pi(u), t(u), \sigma \mapsto \pi(u\sigma)) \in \Delta$.
	The set of runs of $\A$ over $t$ is denoted $\Pi_t(\T)$.
	A run $\pi$ is accepting if, for all $w\in\Sigma^\omega$, the maximal priority that appears infinitely often along the branch $t(w)$, namely $\limsup_{i\to\infty} \theta(\pi(\sigma_0\dots\sigma_i))$, is even.
	The language of $\A$ is $T(\A)=\{t\in\Lambda^\omega_{\Sigma} \st \pi\in\Pi_t(\T), \limsup_{i\to\infty} \theta(\pi(\sigma_0\dots\sigma_i)) \equiv 0 \mod 2\}$, i.e., the set of all trees that admit an accepting run.

	Parity tree automata are expressive enough to recognize the language of tree-encoded assignments for a given resolver logic formula. 
	We describe an automaton with three computational phases.
	In first phase, the automaton guesses its initial state.
	All states hold a vector $\vec{z}$ of $m^2$ weights appearing in $\A_1, \dots, \A_n$ and satisfying $\varphi$, i.e., such that $\varphi[\forall v_{(x, x')}\in V : v_{(x, x')} \gets \vec{z}[x][x']]$ is true.
	Such a vector is guessed at the root of the run tree and carried in all nodes thanks to the states.
	Since there are finitely many weights and free variables in $\varphi$, there are also finitely many vectors $\vec{z}$.
	In the second phase, the automaton ``waits'' for finitely many transitions.
	This is important for $\LimInf$ and $\LimSup$ automata, because the run of $\A_k^{\alpha_t(x_i)}$ over $\alpha_t(x_j)$ may visit finitely many times some weights independent of the long-run value.
	In the third phase, the automaton checks whether the guessed vector $\vec{z}$ is coherent with the tree-encoded assignment that it reads.
	Given a run $\pi$ over the tree $t\in\Lambda^\omega_\Sigma$ that carries the vector $\vec{z}$, for all word variable $x_j\in W$ and all resolver variable $x_i\in F_k$, the value $\vec{z}[x_i][x_j]$ is a coherent assignment for the variable $v_{(x_i, x_j)}\in V$ when $\vec{z}[x_i][x_j] = \A_k^{\alpha_t(x_i)}(\alpha_t(x_j))$.
		Consider the assignment $\alpha_t$ given as an input tree $t\in \Lambda^\omega_{\Sigma}$. 
		The run produced by the resolver $\alpha_t(x_i) \in R(\A_k)$ over the word $\alpha_t(x_j)\in\Sigma^\omega$ corresponds, by construction, 
		to the branch $t_i(\alpha_t(x_j))$.
		To check $\vec{z}[x_i][x_j] = \A_k^{\alpha_t(x_i)}(\alpha_t(x_j))$ at runtime, the automaton ensures that:
		(1) the weight $\vec{z}[x_i][x_j]$ is visited infinitely often along the run $t_i(\alpha_t(x_j))$, 
		and (2) it is never dismissed by another weight (e.g., for $\LimInf$ value function, the guessed weight should be the smallest visited after the waiting phase).
	Hence, the accepting condition of the automaton is such that, for all $x_j \in W$ and all $x_i\in F_k$, if the automaton accepts $t\in\Lambda^\omega_\Sigma$ then $\A_k^{\alpha_t(x_i)}(\alpha_t(x_j))=\vec{z}[x_i][x_j]$.
	Next, we describe how the quantifiers of a resolver logic formula are handled base on the automaton corresponding to its inner Presburger formula.
	
	\smallskip\noindent
	{\color{lipicsGray}\bfseries Handling Quantifiers.~}
	To decide the model-checking of a resolver logic formula of fixed size, we first construct a parity tree automata as presented above.
	Its size is at most $\mathcal{O}(|\max_{1\leq i \leq n} |\A_i|)^{\mathcal{O}(m^2+n)}$.
	Since, the satisfiability of an existential Presburger formula with a fixed number of quantifiers is in \PTime~\cite{ac165798-766d-3e3a-a2ea-7fe0e314d50a}, the automaton can be constructed in polynomial time when $|\varPsi|$ is fixed (i.e., $n$, $m$ and $|\varphi|$ are fixed).
	Then, we release the existentially quantified variables through projections, i.e., leaving the automaton a non-deterministic choice while relaxing a dimension of the input tree.
	Universal quantifiers $\forall x : \varPsi'$ are treated as $\lnot \exists x : \lnot\varPsi'$, where each negation $\lnot$ requires the complementation of the current tree automaton, and then induce an exponential blow up of the computation time~\cite{DBLP:books/ems/21/Loding21}.
	Ultimately, we obtain a tree automaton that does not read labels and the model-checking of the resolver logic formula reduces to its language non-emptiness.
	It is worth emphasizing that, when a universal quantifier appears at the edge of the quantifier sequence, some complementations can be avoided (e.g., when all quantifiers are universal).
	When the innermost quantifier is universal, the automaton is constructed over $\lnot \varphi$ instead of $\varphi$.
	When the outermost quantifier is universal, we leverage the parity acceptance condition of the tree automata to perform a final complementation in $\PTime$.
	Formally, we increase all state priority by 1 and we check the language emptiness instead of non-emptiness.
	
	\begin{remark}
			Resolver logic, as presented above, quantifies over non-partial resolvers (called \emph{full} here to improve clarity).
			We can extend it to handle partial resolvers over product automata.
			The key observation is that, as long as a collection of partial resolvers is conclusive, they collectively define a full resolver.
			Moreover, partial resolvers over the components of product automata are conclusive by definition (see \cref{rm:productresolvers}).
			Hence, the parity tree automaton in the proof of \cref{cl:algo} is constructed similarly.
			Some modifications are necessary to reason about the full resolver defined by the conclusive collection of partial resolvers, but the size of the parity tree automaton constructed above and the overall complexity will not change.
			This is because the partial resolvers are defined over a product of automata, which is already taken into account for full resolvers in the construction above.
	\end{remark}

	To conclude, let us note that our construction allows checking inclusion or simulation at a high cost.
	For example, although checking the inclusion of two $\LimSup$-automata is $\PSpaceC$, using an equivalent formulation from \cref{cl:inclusion}, we obtain a $2\text{-}\ExpTime$ algorithm using the construction presented in this section.
	Similarly, while checking simulation for $\LimSup$-automata can be done in $\NPTime \cap \coNPTime$, we obtain a $3\text{-}\ExpTime$ algorithm using \cref{cl:simulation}.

	\section{Applications of Resolver Logic} \label{sec:applications}
	
	In this section, we explore various applications of resolver logic, highlighting its versatility in addressing problems in automata theory and system verification.
	We begin by presenting how resolver logic can be used for checking strategic dominance and other relations we studied in \cref{sec:resolverrelations}.
	Next, we examine its role in checking the bottom value of automata, which is a crucial problem for deciding co-safety and co-liveness of automata and was left open in \cite{DBLP:conf/concur/BokerHMS23}.
	Then, we explore its application for checking history-determinism of automata, and finally discuss its relevance in checking hyperproperty inclusion.
	As in the previous sections, we note that the results of this section also hold when we only consider boolean safety, reachability, Büchi, and coBüchi automata.
	
	\subsection{Checking Strategic Dominance and Other Resolver-Based Relations} \label{sec:stratdom}

	Let $\A$ and $\B$ be two automata and recall that $\B$ strategically dominates $\A$, denoted $\A \commitsim \B$, iff for all resolvers $f \in R(\A)$ there exists a resolver $g \in R(\B)$ such that $\A^f(w) \leq \B^g(w)$ for all words $w \in \Sigma^\omega$.
	Formulating this condition as a resolver logic formula, we obtain a 2-\ExpTime algorithm for checking strategic dominance thanks to \cref{cl:algo}.
		
	\begin{corollary}
		Let $\nu_1, \nu_2 \in \{\Inf, \Sup, \LimSup, \LimInf\}$.
		For all $\nu_1$-automata $\A$ and all $\nu_2$-automata $\B$, checking whether $\A \commitsim \B$ can be done in 2-\ExpTime.
	\end{corollary}

	Note that other relations introduced in \cref{sec:resolverrelations} can be also checked similarly.
	
	\subsection{Checking the Bottom Value of Automata} \label{sec:bottom}
	
	Safety and liveness \cite{DBLP:journals/ipl/AlpernS85,DBLP:journals/dc/AlpernS87}, as well as co-safety and co-liveness, are fundamental concepts in specification of system properties and their verification.
	These concepts have been recently extended to quantitative properties \cite{DBLP:conf/fossacs/HenzingerMS23}, and safety and liveness have been studied in the context of quantitative automata \cite{DBLP:conf/concur/BokerHMS23}.
	Note that quantitative automata resolve nondeterminism by $\sup$, i.e., given an automaton $\A$ and a word $w$, we have $\A(w) = \A_{\sup}(w)$.
	
	For deciding the safety or liveness of a given automaton $\A$, computing its \emph{top value}, namely the value of $\top_{\A} = \sup_{w \in \Sigma^\omega} \sup_{f \in R(\A)} \A^f(w)$, is shown to be a central step.
	The $\PTime$ algorithm provided in~\cite{DBLP:conf/concur/BokerHMS23} is used as a subroutine for computing the safety closure of a given automaton $\A$, which is used for checking both safety and liveness of $\A$.
	In particular, $\A$ is safe iff the safety closure of $\A$ maps every word $w$ to the same value as $\A$, and $\A$ is live iff the safety closure of $\A$ maps every word $w$ to $\top_{\A}$.
	
	Given an automaton $\A$, we can solve the top-value problem by simply iterating over its weights in decreasing order and checking for each weight $k$ whether there exist $f \in R(\A)$ and $w \in \Sigma^\omega$ with $\A^f(w) \geq k$.
	The largest $k$ for which this holds is the top value of $\A$.
	Note that thanks to \cref{cl:algo} we can achieve this in \PTime as it is only an existential formula, which gives us a new algorithm for this problem.

	However, the problems of deciding the co-safety and co-liveness of automata were left open.
	For these, one needs to compute \emph{bottom value} of a given automaton $\A$, namely $\bot_{\A} = \inf_{w \in \Sigma^\omega} \sup_{f \in R(\A)} \A^f(w)$.
	Similarly as above, using the computation of the bottom value of $\A$ as a subroutine, we can decide its co-safety and co-liveness:
	$\A$ is co-safe iff the co-safety closure of $\A$ maps every word $w$ to the same value as $\A$, and $\A$ is co-live iff the co-safety closure of $\A$ maps every word $w$ to $\bot_{\A}$.
	
	For the classes of automata we consider, we can compute the bottom value in $\PSpace$ by repeated universality checks over its finite set of weights:
	the largest weight $k$ for which the automaton is universal is its bottom value.
	We remark that the bottom value of limit-average automata is uncomputable since their universality is undecidable~\cite{DBLP:conf/csl/DegorreDGRT10,DBLP:conf/concur/ChatterjeeDEHR10}.

	Together with \cref{cl:algo}, the theorem below provides us with a $2\text{-}\ExpTime$ algorithm for computing the bottom value of $\Inf$-, $\Sup$-, $\LimInf$-, and $\LimSup$-automata.
	
	\begin{theorem} \label{cl:bottom}
		Let $\nu \in \{\Inf, \Sup, \LimSup, \LimInf\}$.
		Let $\A$ be a $\nu$-automaton and $x$ be an integer.
		Then, the bottom value of $\A$ is $x$ iff
		$ \exists w_1 \in \Sigma^\omega : \exists f_1 \in R(\A) : \forall w_2 \in \Sigma^\omega : \forall f_2 \in R(\A) : \exists f_3 \in R(\A) : \A^{f_1}(w_1) = x \land \A^{f_1}(w_1) \geq \A^{f_2}(w_1) \land \A^{f_1}(w_1) \leq  \A^{f_3}(w_2)$.
		Moreover, given $\A$ and $x$, this can be checked in $2\text{-}\ExpTime$.
	\end{theorem}
		
	\subsection{Checking History-Determinism of Automata} \label{sec:hdness}

	History-determinism \cite{DBLP:conf/csl/HenzingerP06,DBLP:conf/icalp/Colcombet09} lies between determinism and nondeterminism.
	Intuitively, an automaton is history-deterministic if there exists a way of resolving its nondeterminism based on the current execution prefix (i.e., only the past) while ensuring that the value of the resulting run equals the value assigned to the word by resolving its nondeterminism by $\sup$.
	Although the concept of history-determinism first appeared as ``good-for-gameness'' in \cite{DBLP:conf/csl/HenzingerP06}, following the distinction made in \cite{DBLP:conf/fsttcs/BokerL21}, we use the definition of history-determinism in \cite{DBLP:conf/icalp/Colcombet09}.

	\begin{definition}[history-determinism \cite{DBLP:conf/icalp/Colcombet09}]
		Let $\A = (\Sigma, Q, s, \Delta, \mu, \nu)$ be an automaton.
		Then, $\A$ is \emph{history-deterministic} iff Player-2 wins the letter game defined below.
		
		The \emph{letter game} on $\A$ is played as follows:
		The game begins on the initial state $q_0 = s$.
		At each turn $i \geq 0$ that starts in a state $q_i$, Player-1 first chooses a letter $\sigma_i \in \Sigma$, then Player-2 chooses a transition $d_i = (q_i, \sigma_i, q_{i+1}) \in \Delta$, and the game proceeds to state $q_{i+1}$.
		The corresponding infinite play is an infinite run $\pi$ over the word $w = \sigma_0 \sigma_1 \ldots$, and Player-2 wins the game iff $\A_{\sup}(w) \leq \nu(\mu(\pi))$.
	\end{definition}

	\begin{remark} \label{rm:hdness}
		An automaton $\A$ is history-deterministic iff there exists a resolver $f \in R(\A)$ such that $\A_{\sup}(w) \leq \A^f(w)$ for all $w \in \Sigma^\omega$.
		One can verify that the resolver $f$ is exactly the winning strategy for Player-2 in the letter game on $\A$.
	\end{remark}

	History-deterministic automata offer a balance between deterministic and nondeterministic counterparts, with notable advantages.
	For instance, history-deterministic $\LimInf$-automata are exponentially more concise than deterministic ones \cite{DBLP:conf/icalp/KuperbergS15}, and history-deterministic pushdown automata exhibit both increased expressiveness and at least exponential succinctness compared to their deterministic counterparts \cite{DBLP:journals/lmcs/GuhaJLZ24}.
	Further exploration is detailed in \cite{DBLP:journals/siglog/BokerL23,DBLP:journals/lmcs/BokerL23}.
	
	In \cite{DBLP:journals/lmcs/BokerL23}, algorithms are presented to determine whether an automaton is history-deterministic.
	The approach involves solving a token game that characterizes history-determinism for the given automata type.
	The procedure is in $\PTime$ for $\Inf$- and $\Sup$-automata, quasipolynomial time for $\LimSup$, and $\ExpTime$ for $\LimInf$.
	Combined with \cref{cl:algo}, the theorem below presents a new $\ExpTime$ algorithm for checking history-determinism across all these automata types, providing competitive complexity with \cite{DBLP:journals/lmcs/BokerL23} for $\LimInf$-automata.
	
	\begin{theorem} \label{cl:hdness}
		Let $\nu \in \{\Inf, \Sup, \LimSup, \LimInf\}$ and $\A$ be a $\nu$-automaton.
		Then, $\A$ is history-deterministic iff $\A \blindsim \A$.
		Moreover, given $\A$, this can be checked in $\ExpTime$.
	\end{theorem}
	
	\subsection{Checking Hyperproperty Inclusion} \label{sec:hyper}

	We have focused on trace properties---functions mapping words to values, either 0 or 1 in the boolean setting.
	While adept at representing temporal event orderings, trace properties lack the capacity to capture dependencies among multiple system executions, such as noninterference in security policies \cite{DBLP:conf/sp/GoguenM82a} or fairness conditions for learning-based systems \cite{DBLP:conf/atva/SeshiaDDFGKSVY18}.
	
	This limitation is addressed by hyperproperties \cite{DBLP:journals/jcs/ClarksonS10}.
	Unlike trace properties, hyperproperties encompass global characteristics applicable to sets of traces.
	This enables the specification of intricate relationships and constraints beyond temporal sequencing.
	Formally, while a trace property is a set of traces, a hyperproperty is a set of trace properties.
	
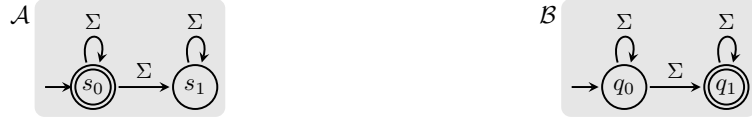
\begin{figure}[t]\centering
		\noindent\begin{minipage}{.49\linewidth}\centering
			\scalebox{\picturescaletikz}{
			\begin{tikzpicture}[bg={\A}, node distance=.8cm]
				\node [state, final, initial, label=center:$s_0$] (q0) {};
				\node [right=of q0, state, label=center:$s_1$] (q1) {};
				\path[transition]
				(q0) edge[loop above] node {$\Sigma$} (q0)
				(q1) edge[loop above] node {$\Sigma$} (q1)
				(q0) edge node[above]{$\Sigma$} (q1)
				;
			\end{tikzpicture}
			}
		\end{minipage}
		\noindent\begin{minipage}{.49\linewidth}\centering
			\scalebox{\picturescaletikz}{
				\begin{tikzpicture}[bg={\B}, node distance=.8cm]
					\node [state, initial, label=center:$q_0$] (q0) {};
					\node[state, right=of q0, label=center:$q_1$, final] (q1) {};
					\path[transition]
					(q0) edge[loop above] node {$\Sigma$} (q0)
					(q1) edge[loop above] node {$\Sigma$} (q1)
					(q0) edge node[above]{$\Sigma$} (q1)
					;
			\end{tikzpicture}
			}
		\end{minipage}
		\caption{Two nondeterministic automata $\A$ and $\B$ over a finite alphabet $\Sigma$ that respectively specify the hyperproperties $\textsf{SP} = \{ P \subseteq \Sigma^\omega \st \text{$P$ is safe} \}$ and $\textsf{CP} = \{ P \subseteq \Sigma^\omega \st \text{$P$ is co-safe} \}$.}\label{fig:hyper}
	\end{figure}

	In this subsection, we use nondeterministic automata as a specification language for hyperproperties.
	A deterministic automaton defines a trace property where each word has a single run, yielding a unique value.
	In contrast, a nondeterministic automaton specifies a trace property only when equipped with a resolver, representing a function from its resolvers to trace properties.
	Formally, a nondeterministic automaton $\A$ specifies the hyperproperty $H_{\A} = \{\A^f \st f \in R(\A)\}$. An illustrative example is presented in \cref{fig:hyper} and \cref{cl:hyperpositive}.
	
	\begin{proposition} \label{cl:hyperpositive}
		The nondeterministic automata $\A$ and $\B$ in \cref{fig:hyper} respectively specify the hyperproperties $\textsf{SP} = \{ P \subseteq \Sigma^\omega \st \text{$P$ is safe} \}$ and $\textsf{CP} = \{ P \subseteq \Sigma^\omega \st \text{$P$ is co-safe} \}$.
	\end{proposition}

	HyperLTL \cite{DBLP:conf/post/ClarksonFKMRS14} extends linear temporal logic (LTL) only with quantification over traces, and therefore cannot express the hyperproperty $\textsf{SP}$ specifying the set of all safety trace properties.
	However, using HyperLTL over the alphabet $\Sigma = \{i, s, o, x\}$, one can express the noninterference between a secret input $s$ and a public output $o$ as follows:
	$\forall \pi, \pi' : \Box(i_{\pi} \leftrightarrow i_{\pi'}) \rightarrow \Box(o_{\pi} \leftrightarrow o_{\pi'})$, i.e., for every pair $\pi, \pi'$ of traces, if the positions of the public input $i$ coincide in $\pi$ and $\pi'$, then so do the positions of the public output $o$.
	We show below that a simpler variant of this property cannot be specified by nondeterministic automata, separating them as a specification language for hyperproperties from HyperLTL. 
	
	\begin{proposition} \label{cl:hypernegative}
		Let $\Sigma = \{a,b,c\}$ and let $\phi = \forall \pi,\pi' : \Box(b_{\pi} \leftrightarrow b_{\pi'})$ be a HyperLTL formula.
		Neither $H_1 = \{ P \subseteq \Sigma^\omega \st \text{$P$ satisfies $\phi$} \}$ nor $H_2 = \{ P \subseteq \Sigma^\omega \st \text{$P$ satisfies $\lnot\phi$} \}$ is expressible by an automaton.
	\end{proposition}

	Together with \cref{cl:algo}, the following theorem gives us a $2\text{-}\ExpTime$ algorithm for checking if the hyperproperty specified by an automaton is included in another one.

	\begin{theorem} \label{cl:hyperinclusion}
		Let $\A$ and $\B$ be two nondeterministic automata respectively denoting the hyperproperties $H_{\A}$ and $H_{\B}$. Then, $H_{\A} \subseteq H_{\B}$ iff $\forall f \in R(\A) : \exists g \in R(\B) : \forall w \in \Sigma^\omega: \A^{f}(w) = \B^{g}(w)$.
		Moreover, given $\A$ and $\B$, this can be checked in 2-$\ExpTime$.
	\end{theorem}

	Using $\A$ for a deterministic automaton representing a system and $\B$ for a nondeterministic automaton defining a hyperproperty, we solve the model checking problem by determining if there exists $g \in R(\B)$ such that $\A(w) = \B^{g}(w)$ for all $w \in \Sigma^\omega$.
	This is solvable in $\ExpTime$ (\cref{cl:algo}).
	Notably, for HyperLTL, which is incomparable to nondeterministic automata as a specification language, the complexity is $\PSpaceH$ in the system's size \cite{DBLP:conf/post/ClarksonFKMRS14}.
	
	
	\section{Conclusion}
	
	We introduced a novel perspective on model refinement, termed \emph{strategic dominance}.
	This view, which falls between trace inclusion and tree inclusion, captures the relationship between two nondeterministic state-transition models by emphasizing the ability of the less precise model to accommodate all deterministic implementations of the more refined one.
	We formally defined strategic dominance and showed that it can be checked in 2-\ExpTime using \emph{resolver logic}---a decidable extension of Presburger logic we developed in this work.
	Resolver logic is a powerful tool for reasoning about nondeterministic boolean and quantitative finite automata over infinite words.
	We provided a model-checking algorithm for resolver logic which, besides the verification of resolver-based refinement relations such as strategic dominance,
	allows the checking of co-safety, co-liveness, and history-determinism of quantitative automata, and the inclusion of hyperproperties specified by nondeterministic automata.
	There are some problems we have left open, including the study of resolver logic for other value functions as well as lower bounds for the model-checking problem of resolver logic and its fragments.
	Future research should also extend resolver logic and its model-checking algorithm to handle the settings of partial information, of multiple agents, and of probabilistic strategies.

	\bibliography{preorders}
	
	\newpage
	\section*{Appendix: Omitted Proofs}

\subsubsection*{Proofs of \cref{sec:resolverrelations}}

\begin{proof}[Proof of \cref{cl:supsupsup}]
	Let $\A = (\Sigma, Q, s, \Delta, \mu, \nu)$.
	Recall that $\nu \in \{\Inf, \Sup, \LimInf, \LimSup\}$ and $\Delta$ is finite.
	Let $w \in \Sigma^\omega$ be a word.
	Since $\A$ has finitely many transitions and thus finitely many weights appearing on them, the set $V(\A, w) = \{\nu(\mu(\pi)) \st \text{$\pi$ is a run of $\A$ on $w$}\}$ of its values for $w$ is also finite.
	Evidently, each run of $\A$ on $w$ is induced by a resolver of $\A$ and vice versa, therefore $V(\A, w) = \{A^f(w) \st f \in R(\A)\}$.
	Due to finiteness of $V(\A, w)$, we get $\sup(V(\A, w)) \in V(\A, w)$.
	Then, there exists a run $\pi$ of $\A$ on $w$ with $\nu(\mu(\pi)) = \sup(V(\A, w))$, and thus a resolver $f \in R(\A)$ such that $\A^{f}(w) = \sup(V(\A, w))$.
\end{proof}

\begin{proof}[Proof of \cref{cl:inclusion}]
	Let $\A$ and $\B$ be two automata.
	We show that $$\A \weakblindsim \B \xRightarrow{(1)} \A \weakblindsim_{\times} \B \xRightarrow{(2)} \A \weakcommitsim_{\times} \B \xRightarrow{(3)} \A \weakcommitsim \B \xRightarrow{(4)} \A \inclusion \B \xRightarrow{(5)} \A \weakblindsim \B.$$
	\begin{enumerate}
		\item
		Assume $\A \weakblindsim \B$, i.e., for all $w \in \Sigma^\omega$ there exists $g \in R(\B)$ such that for all $f \in R(\A)$ we have $\A^f(w) \leq \B^g(w)$.
		Let $w \in \Sigma^\omega$ be arbitrary.
		Notice that the resolver of $\B$ cannot depend on the resolver of $\A$, and it cannot observe the states of $\A$.
		Intuitively, if $\B$ can achieve under this restriction a value greater than $\A^f(w)$ for all $f \in R(\A)$, then it can also achieve this without the restriction, i.e., when the two resolvers operate on the product automaton where they can observe each others' states.
		Now, let $g \in R(\B)$ be as above, and let $g' \in R_2(\A, \B)$ be a partial resolver of $\A \times_2 \B$ that imitates $g$, i.e., $g'(h, \sigma) =\{(q, g(h', \sigma)) \st q \in Q_{\A}\}$ for all $\sigma \in \Sigma$ and every history $h$ over $\A \times_2 \B$ where $h'$ is its projection to its $\B$-component.
		We claim that $(\A \times_1 \B)^{\{f', g'\}}(w) \leq (\A \times_2 \B)^{\{f', g'\}}(w)$ for all $f' \in R_1(\A, \B)$.
		Suppose towards contradiction that for some $f' \in R_1(\A, \B)$ we have $(\A \times_1 \B)^{\{f', g'\}}(w) > (\A \times_2 \B)^{\{f', g'\}}(w)$.
		Consider a resolver $f'' \in R(\A)$ such that $f''(h)$ is the projection of $f'(h)$ on its $\A$-component for any history $h$ over $\A \times_1 \B$.
		Then, we have $\A^{f''}(w) > \B^g(w)$ where $g$ is as above, which contradicts $\A \weakblindsim \B$, therefore $\A \weakblindsim_{\times} \B$.
		
		\item
		Follows from the fact that $\exists x : \forall y : \varphi(x,y)$ implies $\forall y : \exists x : \varphi(x,y)$ in general.
		
		\item 
		The idea is the same as in the proof of $\A \weakblindsim \B \Rightarrow \A \weakblindsim_{\times} \B$.
		Intuitively, if $\B$ can achieve a greater value while $\A$ can base its choices on the choices of $\B$, then it can also achieve this when the choices of $\A$ do not depend on $\B$.
		
		\item 
		Assume $\A \weakcommitsim \B$, i.e., for all $w \in \Sigma^\omega$ and all $f \in R(\A)$ there exists $g \in R(\B)$ such that $\A^f(w) \leq \B^g(w)$.
		Let $w \in \Sigma^\omega$ be arbitrary.
		If for all $f \in R(\A)$ there exists $g \in R(\B)$ such that $\A^f(w) \leq \B^g(w) $, then for all $f \in R(\A)$ we have $\A^f(w) \leq \sup_{g \in R(\B)} \B^g(w)$.
		Since, by \cref{cl:supsupsup}, the value $\sup_{f\in R(\A)} \A^f(w)$ is achieved by some resolver of $\A$, it implies that $\sup_{f\in R(\A)} \A^f(w) \leq \sup_{g \in R(\B)} \B^g(w)$, and thus $\A \inclusion \B$.
		
		\item 
		Assume $\A \inclusion \B$, i.e., $\A_{\sup}(w) \leq \B_{\sup}(w)$ for all $w \in \Sigma^\omega$.
		Let $w \in \Sigma^\omega$ be arbitrary.
		Note that for all $f \in R(\A)$ we have $\A^f(w) \leq \A_{\sup}(w)$, and thus for all $f \in R(\A)$ it holds that $\A^f(w) \leq \B_{\sup}(w)$.
		Since, by \cref{cl:supsupsup}, there exists $g \in R(\B)$ such that $\B_{\sup}(w) = \B^g(w)$, we get that for all $f \in R(\A)$ there exists $g \in R(\B)$ such that $\A^f(w) \leq \B^g(w)$.
	\end{enumerate}
\end{proof}

\begin{proof}[Proof of \cref{cl:simulation}]
	Let $\A = (\Sigma, Q_{\A}, s_{\A}, \Delta_{\A}, \mu_{\A}, \nu_{\A})$ and $\B = (\Sigma, Q_{\B}, s_{\B}, \Delta_{\B}, \mu_{\B}, \nu_{\B})$.
	
	$\A \commitsim_{\times} \B \Rightarrow \A \simulation \B$:
	We prove the contrapositive.
	Assume $\A \not \simulation \B$, i.e., Challenger has a strategy $\tau$ such that for all outcomes $(r_1, r_2)$ of $\tau$ we have $\nu_{\A}(\mu_{\A}(r_1)) > \nu_{\B}(\mu_{\B}(r_2))$.
	We want to show that there exists $f \in R_1(\A, \B)$ such that for all $g \in R_2(\A, \B)$ there exists $w \in \Sigma^\omega$ such that $(\A\times_1\B)^{\{f, g\}}(w) > (\A\times_2\B)^{\{f, g\}}(w)$.
	Let $\tau$ be the winning strategy for Challenger.
	Let $f$ be such that for every history $h = (s_0,t_0) a_1 (s_1,t_1) a_2 \ldots (s_n,t_n) \in \Pi_{\A \times_1 \B}$ (which is also a history of $\A \times_2 \B$) and letter $\sigma \in \Sigma$, if $\tau((s_0,t_0)(s_1,t_1) \ldots (s_n,t_n)) = (\sigma, q)$ for some $q \in Q_{\A}$, then $f(h, \sigma) = \{(q,p) \st p \in Q_{\B}\}$.
	Now, let $g \in R_2(\A, \B)$ be arbitrary.
	Consider an outcome $(r_1,r_2)$ of $\tau$ such that $r_1 = q_0 \sigma_1 q_1 \sigma_2 \ldots$ and $r_2 = p_0 \sigma_1 p_1 \sigma_2 \ldots$ satisfy $f((q_0, p_0) \sigma_1 \ldots (q_i, p_i), \sigma_{i+1}) \cap g((q_0, p_0) \sigma_1 \ldots (q_i, p_i), \sigma_{i+1}) = \{(q_{i+1}, p_{i+1})\}$ for all $i \geq 0$, which exist by construction.
	Note that $\nu_{\A}(\mu_{\A}(r_1)) > \nu_{\B}(\mu_{\B}(r_2))$ since $\tau$ is winning.
	Then, letting $w = \sigma_1 \sigma_2 \ldots$ gives us $(\A\times_1\B)^{\{f, g\}}(w) > (\A\times_2\B)^{\{f, g\}}(w)$.
	
	$\A \simulation \B \Rightarrow \A \commitsim_{\times} \B$:
	We prove the contrapositive.
	Assume $\A \not \commitsim_{\times} \B$, i.e., there exists $f \in R_1(\A, \B)$ such that for all $g \in R_2(\A, \B)$ there exists $w \in \Sigma^\omega$ such that $(\A\times_1\B)^{\{f, g\}}(w) > (\A\times_2\B)^{\{f, g\}}(w)$.
	We want to show that there exists a strategy $\tau$ for Challenger such that for all outcomes $(r_1, r_2)$ of $\tau$ we have $\nu_{\A}(\mu_{\A}(r_1)) > \nu_{\B}(\mu_{\B}(r_2))$.		
	Let $f \in R_1(\A, \B)$ be a partial resolver as above.
	Let $g \in R_2(\A, \B)$ be arbitrary, and let $w = \sigma_1 \sigma_2 \ldots$ such that $(\A\times_1\B)^{\{f, g\}}(w) > (\A\times_2\B)^{\{f, g\}}(w)$.
	First, let $q_0 = s_{\A}$ and $p_0 = s_{\B}$, and define $\tau((q_0,p_0)) = (\sigma_1, q_1)$ where $f((q_0,p_0), \sigma_1) \cap g((q_0,p_0), \sigma_1) = \{(q_1, p_1)\}$.
	Now, for all $i \geq 1$, define $\tau((q_0,p_0) \ldots (q_i,p_i)) = (\sigma_{i+1}, q_{i+1})$ where $f((q_0,p_0) \sigma_1 \ldots (q_i,p_i), \sigma_{i+1}) \cap g((q_0,p_0) \sigma_1 \ldots (q_i,p_i), \sigma_{i+1}) = \{(q_{i+1}, p_{i+1})\}$.
	Since we chose $g$ to be arbitrary, $\tau$ is defined for every valid history $h$ of the product automaton such that the corresponding word $w$ is a witness to the initial assumption $(\A\times_1\B)^{\{f, g\}}(w) > (\A\times_2\B)^{\{f, g\}}(w)$.
	In the same vein, $(q_0 \sigma_1 q_1 \sigma_2 \ldots, p_0 \sigma_1, p_1 \sigma_2 \ldots)$ is an arbitrary outcome of $\tau$, and since $(\A\times_1\B)^{\{f, g\}}(w) > (\A\times_2\B)^{\{f, g\}}(w)$ where $w = \sigma_1 \sigma_2 \ldots$, we obtain $\nu_{\A}(\mu_{\A}(r_1)) > \nu_{\B}(\mu_{\B}(r_2))$ as required.
	
	An alternative proof of this direction can be given by extending the tree-containment view of fair simulation \cite{DBLP:journals/iandc/HenzingerKR02} to the quantitative automata types we consider.
	From this extension, we would obtain that $\A \simulation \B$ iff $\B$ tree-contains $\A$ in the quantitative sense, i.e., for every run-tree $T$ of $\A$ there is a run-tree $T'$ of $\B$ such that $T$ and $T'$ are witnessed by the same trace-tree and for each path $\pi$ in $T$ the corresponding path $\pi'$ in $T'$ satisfies $\nu_{\A}(\mu_{\A}(\pi)) \leq \nu_{\B}(\mu_{\B}(\pi'))$. 
	Then, to prove $\A \commitsim_{\times} \B$, we only need to argue that for each deterministic run-tree of $\A$ there is a deterministic run-tree of $\B$ satisfying the above condition, which holds by definition of simulation.
\end{proof}

\begin{proof}[Proof of \cref{cl:implications}]
	Let $\A$ and $\B$ be two automata.
	We prove each statement separately.
	\begin{enumerate}
		\item
		Assume $\A \commitsim \B$, i.e., for all $f \in R(\A)$ there exists $g \in R(\B)$ such that for all $w \in \Sigma^\omega$ we have $\A^f(w) \leq \B^g(w)$.
		Equivalently, for all $f \in R(\A)$ there exists $g \in R(\B)$ such that $\inf_{w \in \Sigma^\omega} (\B^g(w) - \A^f(w)) \geq 0$.
		This is equivalent to the following: for all $f \in R(\A)$ we have $\sup_{g \in R(\B)} \inf_{w \in \Sigma^\omega} (\B^g(w) - \A^f(w)) \geq 0$.
		Finally, another equivalent formulation is $\inf_{f \in R(\A)} \sup_{g \in R(\B)} \inf_{w \in \Sigma^\omega} (\B^g(w) - \A^f(w)) \geq 0$.
		Now, observe that this implies $\inf_{f \in R(\A)} \inf_{w \in \Sigma^\omega} \sup_{g \in R(\B)} (\B^g(w) - \A^f(w)) \geq 0$ (but not vice versa), which is equivalent to $\inf_{w \in \Sigma^\omega} \inf_{f \in R(\A)} \sup_{g \in R(\B)} (\B^g(w) - \A^f(w)) \geq 0$.
		From this, we obtain first $\inf_{w \in \Sigma^\omega} \inf_{f \in R(\A)} (\B_{\sup}(w) - \A^f(w)) \geq 0$, and then $\inf_{w \in \Sigma^\omega} (\B_{\sup}(w) -  \A_{\sup}(w)) \geq 0$.
		Finally, this means $(\B_{\sup}(w) -  \A_{\sup}(w)) \geq 0$ for all $w \in \Sigma^\omega$, and thus $\A \inclusion \B$.
		
		\item
		Thanks to \cref{cl:simulation}, it suffices to show that $\A \commitsim_{\times} \B$ implies $\A \commitsim \B$.
		The reasoning is the same as in the proof of $\A \weakcommitsim_{\times} \B \Rightarrow \A \weakcommitsim \B$ (item (3) of \cref{cl:inclusion}).
		Intuitively, since resolvers over synchronized products can observe the other component's history (as opposed to resolvers over single automata), if all partial $\A$-resolvers of $\A \times_1 \B$ fail to achieve a greater value against $\A \times_2 \B$, then so do all resolvers of $\A$ against $\B$.
		
		\item
		Thanks to \cref{cl:simulation}, it suffices to show that $\A \blindsim_{\times} \B$ implies $\A \commitsim_{\times} \B$.
		To show this, simply recall the definitions of $\blindsim_{\times}$ and $\commitsim_{\times}$, and note the fact that $\exists x : \forall y : \varphi(x,y)$ implies $\forall y : \exists x : \varphi(x,y)$ in general.
		
		\item
		The reasoning is the same as in the proof of $\A \weakblindsim \B \Rightarrow \A \weakblindsim_{\times} \B$ (item (1) of \cref{cl:inclusion}).
		Intuitively, if there is a resolver of $\B$ that achieves greater value against $\A$ (without being able to observe the history of $\A$), then there is a partial $\B$-resolver of $\A \times_2 \B$ that achieves a greater value against $\A \times_1 \B$ (since it can now observe the history of $\A$).\qedhere
	\end{enumerate}
\end{proof}

\begin{proof}[Proof of \cref{cl:restrictedresolvers}]
	Simply observe that the proofs in \cref{cl:inclusion,cl:simulation,cl:implications} do not depend on the memory capabilities of resolvers.
	Whenever we speak of resolvers or strategies, we now consider only the appropriately restricted ones, and the same arguments apply.
\end{proof}


\begin{proof}[Proof of \cref{cl:counterexamples}]
	We prove each item separately.
	\begin{enumerate}
		\item
		Let $\A$ and $\B$ as in \cref{fig:sep1}.
		Note that $\A$ has only one resolver and $\B$ has four resolvers all of which is positional.
		
		$\A \inclusion^{\text{pos}} \B$:
		One can verify that for all words $\B$ has a run whose value is 1.
		
		$\A \not\commitsim^{\text{pos}} \B$:
		We want to show that there exists $f \in R(\A)$ such that for all $g \in R(\B)$ there exists $w \in \Sigma^\omega$ such that $\A^f(w) > \B^g(w)$.
		Since $\A$ is deterministic and the value domain is binary, it suffices to show for all $g \in R(\B)$ there exists $w \in \Sigma^\omega$ such that $\B^g(w) = 0$.
		Let $g$ be an arbitrary resolver of $\B$ and let $\sigma \in \{a,b\}$.
		If $g$ moves $\B$ from $q_0$ to $q_1$ with $\sigma$, then let $w = \sigma b^\omega$.
		Otherwise, if $g$ moves $\B$ from $q_0$ to $q_2$ with $\sigma$, then let $w = \sigma a^\omega$.
		
		\item
		Let $\A$ and $\B$ as in \cref{fig:sep2sep4}.
		Note that both $\A$ and $\B$ have only two resolvers and all of them are positional.
		
		$\A \commitsim^{\text{pos}} \B$:
		We want to show that for all $f \in R(\A)$ there exists $g \in R(\B)$ such that $\A^f(w) \leq \B^g(w)$ for all $w \in \Sigma^\omega$.
		Let $f$ be an arbitrary resolver of $\A$, which moves $\A$ from $s_1$ to either $s_2$ or $s_3$ with $a$.
		If it moves to $s_2$, then consider a resolver $g$ that moves $\B$ from $q_0$ to $q_1$ with $a$.
		For these, one can verify that $\A^f(w) = \B^g(w)$ for all $w \in \Sigma^\omega$.
		Otherwise, if moves to $s_3$, then consider a resolver $g$ that moves $\B$ to $q_2$.
		One can similarly verify that these satisfy the desired outcome.
		
		$\A \not\simulation^{\text{pos}} \B$:
		It is easy to see that $s_0$ is not simulated by $q_0$.
		
		\item
		Let $\A$ be as in \cref{fig:sep2sep4}.
		Observe that $\A$ has two resolvers, both of which are positional.
		
		$\A \simulation^{\text{pos}} \A$:
		Trivial since the simulation relation is reflexive.
		
		$\A \not\blindsim_{\times}^{\text{pos}} \A$:
		We want to show that for all $g \in R_2(\A, \A)$ there exist $f \in R_1(\A, \A)$ and $w \in \Sigma^\omega$ such that  $(\A \times_1 \A)^{\{f, g\}}(w) > (\A \times_2 \A)^{\{f, g\}}(w)$.
		Note that $\A \times_1 \A = \A \times_2 \A$.
		Let $g \in R_2(\A, \A)$ be an arbitrary partial resolver for the second component of the product.
		If $g$ moves the product from $(s_1, s_1)$ to $\{(q, s_2) \st q \in Q_{\A}\}$, then consider the resolver $f \in R_1(\A, \A)$ that moves the product from $(s_1, s_1)$ to $\{(s_3, q) \st q \in Q_{\A}\}$ and the word $w = ab^\omega$.
		Otherwise, if $g$ moves the product to $\{(q, s_3) \st q \in Q_{\A}\}$, then consider the resolver $f \in R_1(\A, \A)$ that moves it to $\{(s_2, q) \st q \in Q_{\A}\}$ and the word $w = a^\omega$.
		
		\item
		Let $\A$ and $\B$ as in \cref{fig:sep2sep4}.
		Note that both $\A$ and $\B$ have only two resolvers each and all of them are positional.
		
		$\B \blindsim_{\times}^{\text{pos}} \A$:
		We want to show that there exists $g \in R_2(\B, \A)$ such that for all $f \in R_1(\B, \A)$ and all $w \in \Sigma^\omega$ we have $(\B \times_1 \A)^{\{f, g\}}(w) \leq (\B \times_2 \A)^{\{f, g\}}(w)$.
		Let $g$ be a partial resolver that moves the second component of the product with $a$ from $(q_1,s_1)$ to $\{(q, s_2) \st q \in Q_{\B}\}$ and from $(q_2,s_1)$ to $\{(q, s_3) \st q \in Q_{\B}\}$.
		Let $f \in R_1(\B, \A)$ be arbitrary.
		Note that $f$ moves the first component of the product with $a$ from $(q_0, s_0)$ to either $\{(q_1, s) \st s \in Q_{\A}\}$ or $\{(q_2, s) \st s \in Q_{\A}\}$.
		Since $g$ only has one option for this transition, the next state is either $(q_1, s_1)$ or $(q_2, s_1)$, which the two resolvers respectively move with $a$ to $(q_3, s_2)$ or $(q_4, s_3)$.
		One can verify that any infinite suffix after reaching these states leads to the same values on both product automata.
		
		$\B \not\blindsim^{\text{pos}} \A$:
		We want to show that for all $g \in R(\A)$ there exist $f \in R(\B)$ and $w \in \Sigma^\omega$ such that $\B^f(w) > \A^g(w)$.
		Let $g \in R(\A)$ be arbitrary.
		If $g$ moves $\A$ from $s_1$ to $s_2$ with $a$, consider the resolver $f \in R(\B)$ that moves $\B$ from $q_0$ to $q_2$ with $a$ and the word $w = a a b^\omega$.
		Otherwise, if $g$ moves $\A$ from $s_1$ to $s_3$ with $a$, consider the resolver $f \in R(\B)$ that moves $\B$ from $q_0$ to $q_1$ with $a$ and the word $w = a^\omega$.\qedhere
	\end{enumerate}
\end{proof}

\subsubsection*{Proofs of \cref{sec:algorithm}}
\begin{proof}[Proof of \cref{cl:algo}]
	We assume that each automata $\A_k=(Q_k, s_k, \Delta_k, \mu_k, \nu_k)$ is either a $\LimSup$-automaton or a $\LimInf$-automaton.
	This is without loss of generality since $\Sup$-automata and $\Inf$-automata can be converted in \PTime into $\LimInf$-automata~\cite[Proposition 2.1]{DBLP:conf/concur/BokerHMS23}.
	The proof goes as follows.
	First, we construct in polynomial time a parity tree automaton $\C$ which read an assignment $\alpha$ for $\varPsi$ as input, such that its language is empty if and only if $\alpha \models \varPsi$.
	Then we handle the quantifiers of $\varPsi$ based on nested complementations and projections applied on $\C$.
	Finally, we construct a parity game such that the even player wins iff $\varPsi$ is satisfiable over $\A_1, \ldots, \A_n$.\hfill\break

	\noindent{\color{lipicsGray}\sffamily Construction of $\C$.}
	Let $\varPsi$ be of the form $\nabla_1 x_1 : \dots : \nabla_m x_m : \varphi$, where $\nabla_i \in \{\exists, \forall\}$.
	In this construction, we encode assignments for resolver and word variables of $\varPsi$ into single $\Sigma$-structured trees.
	The labeling alphabet is defined from the sets $Q_1, \dots, Q_n$ and $\{0,1\}$ in order to manipulate branches as runs.
	For all $i\in\{1, \dots, m\}$, we define $\Lambda_i$ and $\ell_i\in\Lambda_i$ such that if $x_i\in W$ then $\Lambda_i=\{0,1\}$ and $\ell_i=1$; otherwise, $x_i\in F_k$ for some $k\in\{1, \dots, n\}$, and so $\Lambda_i=Q_k$ and $\ell_i=s_k$ where $s_k$ is the initial state of $\A_k$.
	Let the labeling alphabet be $\Lambda = \Lambda_1 \times \dots \times \Lambda_m$ and let the label of roots be $\ell = \ell_1 \times \dots \times \ell_m$.
	For all $\lambda \in \Lambda$ and $1 \leq i \leq m$ we write $\lambda[i]$ to denote the dimension of $\lambda$ corresponding to $\Lambda_i$. 
	In the same way, we construct the value domains of the variables of $V$ from the sets of weights of $\A_1, \dots, \A_k$.
	For all $x_i\in F_k$ and all $x_j\in W$, we define the value domain of the variable $v_{(x_i, x_j)} \in V$ as $Z_{(x_i,x_j)} = \{\mu_k(\delta)\in \NN \st \delta\in\Delta_k\}$.
	Let $Z = \Pi_{k=1}^n \Pi_{f\in F_k} \Pi_{w\in W} Z_{(f,w)}$ be the set of assignment of the variable of $V$.
	For all $z \in Z$, all $x_i\in F_k$ and all $x_j\in W$, we write $z[x_i][x_j]$ to denote the dimension of $z$ corresponding to $Z_{(f,w)}$.
	
	We now construct the parity tree automaton $\C = (\Lambda, \Sigma, Q, I, \Delta, \theta)$.
	The set of $\heartsuit$-states is $Q_{\heartsuit} = \{(\heartsuit_{(y_1, y_2)}, z, \lambda) \st y_1, y_2 \in \{1, \dots, m\}, z\in Z, \lambda\in\Lambda\}$, the set of $\spadesuit$-states is $Q_{\spadesuit} = \{(\spadesuit_{(y_1, y_2)}, z, \lambda) \st y_1, y_2 \in \{1, \dots, m\}, z\in Z, \lambda\in\Lambda\}$, and the set of $\bot$-states is $Q_\bot = \{(\bot, z, \lambda) \st z\in Z, \lambda\in\Lambda\}$.
	The set of states is $Q = Q_{\bot} \cup Q_{\heartsuit} \cup Q_{\spadesuit}$ and the set of initial states is $I = \{(\bot, z, \ell) \st z\in Z,\varphi[\forall v_{(x, x')}\in V : v_{(x, x')} \gets z[x][x']]\}$.
	The priority function $\theta \colon Q \to \{1, 2\}$ maps $\heartsuit$-states to 2 and all the other states to 1.
	The transition relation $\Delta$ is defined as follows.
	\begin{itemize}
		\item $((\bot, z, \lambda), \lambda, \sigma \mapsto (S_\sigma, z, \lambda_{\sigma})) \in\Delta$ where $S_\sigma \in \{\bot, \spadesuit_{(1, 1)}\}$ iff
		$$\bigwedge\begin{cases}
			\bigwedge_{j=1}^m \Big(x_j \in W \Rightarrow \sum_{\sigma\in\Sigma}\lambda_{\sigma}[j]=\lambda[j]\Big)
			\\
			\bigwedge_{i=1}^m \bigwedge_{k=1}^n \bigwedge_{\sigma\in\Sigma}
			\Big( x_i \in F_k \Rightarrow (\lambda[i], \sigma, \lambda_\sigma[i])\in\Delta_k \Big)
		\end{cases}
		$$
	\end{itemize}
	For all transitions, $\C$ ensures that the encoding of its assignment for $x_1, \dots, x_m$ is a coherent $\Lambda$-labeled $\Sigma$-structured tree.
	Above, the first constraint guarantees that all dimensions encoding a word have exactly one branch labeled by 1 (which encodes the word), and all other nodes are labeled by 0.
	Formally, each tree $t\in T(\C)$ assigns the variable $x_j\in W$ to the unique word $\alpha_t(x_j)=\sigma_1\sigma_2\dots \in\Sigma^\omega$ for which the $j$th dimension of the branch $t(\alpha_t(x_j))$ equals $1 \sigma_1 1 \sigma_2\dots \in (\{1\} \times \Sigma)^\omega$.
	The second constraint guarantees that all dimensions encoding a resolver of $\A_k$ respect its transition relation, i.e., a node labeled by $\lambda$ and its $\sigma$-child labeled by $\lambda_{\sigma}$ must encode a transition of $\A_k$ in these dimensions.
	Formally, each tree $t\in T(\C)$ assigns the variable $x_i\in F_k$ to the unique resolver $\alpha_t(x_i)\in R(\A_k)$ defined by $\alpha_t(x_i)(\pi_k, \sigma) = t(u\sigma)$ where $\pi_k$ is the finite run of $\A_k^{\alpha_t(x_i)}$ over $u\in \Sigma^*$.
	In particular, for all $t\in T(\C)$, all $x_i\in F_k$ and $x_j\in W$, the value $\A_k^{\alpha_t(x_i)}(\alpha_t(x_j))$ is the correct assignment for the free variable $v_{(x_i,x_j)}\in V$ of $\varphi$.
	\begin{itemize}	
		\item $((\spadesuit_{(y_1, y_2)}, z, \lambda), \lambda, \sigma \mapsto (\spadesuit_{(y_1, y_2)}, z, \lambda_{\sigma})) \in\Delta$
		iff
		$$\bigwedge\begin{cases}
			\bigwedge_{j=1}^m \Big(x_j \in W \Rightarrow \sum_{\sigma\in\Sigma}\lambda_{\sigma}[j]=\lambda[j]\Big)
			\\
			\bigwedge_{i=1}^m \bigwedge_{k=1}^n \bigwedge_{\sigma\in\Sigma}
			\Big( x_i \in F_k \Rightarrow (\lambda[i], \sigma, \lambda_\sigma[i])\in\Delta_k \Big)
			\\
			\bigwedge_{i=1}^m \bigwedge_{k=1}^n \bigwedge_{j=1}^m \bigwedge_{\sigma\in\Sigma}
			\Big(\big(x_i \in F_k \land \lambda[j]=1\big) \Rightarrow  h_k(z[x_i][x_j], \mu_k(\lambda[i], \sigma, \lambda_\sigma[i]))=z[x_i][x_j]\Big)
		\end{cases}$$
		where $h_k = \max$ if $\A_k$ is a $\LimSup$-automaton and $h_k = \min$ if it is a $\LimInf$-automaton.
	\end{itemize}
	Observe that $\bot$-states are reachable only from $\bot$-states and cannot lead to acceptance as their priority is odd.
	Once a $\spadesuit_{(1,1)}$-state is reached, $\C$ checks through the rest of the run tree whether $z$ provides a correct assignment for the variable of $V$.
	By construction, $z$ is guessed at the root of the run tree and carried in all its nodes.
	Given a run $\pi$ of $\C$ over $t\in\Lambda^\omega_\Sigma$ that carries $z\in Z$, for all $x_j\in W$, all $x_i\in F_k$, the value $z[x_i][x_j]$ is a correct assignment for $v_{(x_i, x_j)}$ when $z[x_i][x_j] = \A_k^{\alpha_t(x_i)}(\alpha_t(x_j))$.
	Intuitively, $v_{(x_i, x_j)}$ requires $\C$ to ensure that the weight $z[x_i][x_j]$ is ($\dagger$) visited infinitely often, ($\ddagger$) never dismissed by another weight, and so along the branch induced by the word $\alpha_t(x_j)$.
	The condition ($\dagger$) is handled by $\C$ thanks to its acceptance condition that we explain below.
	Above, the last constraint guarantees the condition ($\ddagger$), i.e., assuming that $x_j\in W$, $x_i\in R(\A_k)$ and that the current node belongs to the branch induced by $\alpha_t(x_j)$, if $\A_k$ is a $\LimSup$-automaton then the weight of the transition $(\lambda[i], \sigma, \lambda_{\sigma}[i]) \in \Delta_k$ from the node to its $\sigma$-child is at most $z[x_i][x_j]$, otherwise $\A_k$ is a $\LimInf$-automaton and the weight of this transition is at least $z[x_i][x_j]$.
	This constrain appears in all transitions outgoing from the $\spadesuit$-states and the $\heartsuit$-states.
	\begin{itemize}			
		\item $((\spadesuit_{(y_1, y_2)}, z, \lambda), \lambda, \sigma \mapsto (S_{\sigma(y_1, y_2)}, z, \lambda_{\sigma})) \in\Delta$ where $S_{\sigma(y_1,y_2)} \in\{\heartsuit_{(y_1, y_2)}, \spadesuit_{(y_1,y_2)}\}$ iff
		$$\bigwedge\begin{cases}
			\bigwedge_{j=1}^m \Big(x_j \in W\Rightarrow \sum_{\sigma\in\Sigma}\lambda_{\sigma}[j]=\lambda[j]\Big)
			\\
			\bigwedge_{i=1}^m \bigwedge_{k=1}^n \bigwedge_{\sigma\in\Sigma}
			\Big(x_i \in F_k \Rightarrow (\lambda[i], \sigma, \lambda_\sigma[i])\in\Delta_k \Big)
			\\
			\bigwedge_{i=1}^m \bigwedge_{k=1}^n \bigwedge_{j=1}^m \bigwedge_{\sigma\in\Sigma}
			\Big(\big(x_i \in F_k\land \lambda[j]=1\big) \Rightarrow  h_k(z[x_i][x_j], \mu_k(\lambda[i], \sigma, \lambda_\sigma[i]))=z[x_i][x_j]\Big)
			\\
			\bigwedge_{k=1}^n \bigwedge_{\sigma\in\Sigma}
			\Big( \big(x_{y_1} \in F_k \land \lambda[y_2]=1 \land S_{\sigma(y_1, y_2)} = \heartsuit_{(y_1, y_2)}\big) 
			\Rightarrow \mu_k(\lambda[y_1], \sigma, \lambda_\sigma[y_1])=z[x_{y_1}][x_{y_2}]\Big)
		\end{cases}$$
	\end{itemize}
	Given a run $\pi$ of $\C$ over $t\in\Lambda^\omega_\Sigma$ that carries $z\in Z$, assuming that $x_{y_2}\in W$ and $x_{y_1} \in F_k$, the condition ($\dagger$) asks $\C$ to check whether the guessed value $z[x_{y_1}][x_{y_2}]$ is among the values visited infinitely many times along the branch induced by $\alpha_t(x_{y_2})\in\Sigma^\omega$. 
	Above, the last constraint guarantees that $\C$ allows to move from a $\spadesuit_{(y_1,y_2)}$-state to a $\heartsuit_{(y_1,y_2)}$-state only if either $x_{y_1} \notin R(\A_k)$, or the current node does not belong to the branch induced by $\alpha_t(x_{y_2}) \in \Sigma^\omega$, or the guessed weight $z[x_{y_1}][x_{y_2}]$ is visited in the corresponding dimension.
	Observe that $\heartsuit$-states have priority 2 in $\C$, while $\spadesuit$-states have priority 1.
	The condition ($\dagger$) on $z[x_{y_1}][x_{y_2}]$ holds for all accepting runs because $\C$ ensures that a $\heartsuit_{(y_1,y_2)}$-state is visited infinitely many times on all branches of its accepting runs, as we explain below.
	\begin{itemize}	
		\item $((\heartsuit_{(y_1, y_2)}, z, \lambda), \lambda, \sigma \mapsto (\spadesuit_{(y'_1, y'_2)}, z, \lambda_{\sigma})) \in\Delta$
		iff
		$$\bigwedge\begin{cases}
			\bigwedge_{j=1}^m \Big(x_j\in W\Rightarrow \sum_{\sigma\in\Sigma}\lambda_{\sigma}[j]=\lambda[j]\Big)
			\\
			\bigwedge_{i=1}^m \bigwedge_{k=1}^n \bigwedge_{\sigma\in\Sigma}
			\Big(x_i\in F_k \Rightarrow (\lambda[i], \sigma, \lambda_\sigma[i])\in\Delta_k \Big)
			\\
			\bigwedge_{i=1}^m \bigwedge_{k=1}^n \bigwedge_{j=1}^m \bigwedge_{\sigma\in\Sigma}
			\Big(\big(x_i\in F_k\land \lambda[j]=1\big) \Rightarrow  h_k(z[x_i][x_j], \mu_k(\lambda[i], \sigma, \lambda_\sigma[i]))=z[x_i][x_j]\Big)
			\\
			(y'_1=y_1 \land y'_2=y_2+1) \lor (y'_1=y_1+1\land y_2=m \land y'_2=1) \lor (y_1 = y_2 = m \land y'_1= y'_2 = 1)
		\end{cases}$$
	\end{itemize}
	We recall that $\mathcal{O}(m^2)$ values are checked by $\C$ through its runs.
	To ensure that condition ($\dagger$) holds for all dimensions of $z$, the transitions of $\C$ enforces to visit cyclically all $\heartsuit_{(y_1, y_2)}$-states in order to get a run of even priority.
	Above, the last constraint guarantees that $\C$ allows to leave a $\heartsuit_{(y_1,y_2)}$-state only toward a $\spadesuit$-state that regulates the next pair of index.
	Since there is no transition from a $\heartsuit$-state to a $\heartsuit$-state, a run of $\C$ is accepting if and only if all branches visit a $\heartsuit_{(y_1,y_2)}$-state infinitely often for all $y_1,y_2\in\{1, \dots, m\}$.
	As final observation, we point out that, since $z$ carried in all nodes of the run tree, the consistency of ($\dagger$) and ($\ddagger$) through branches is guaranteed.
	Hence, for all $t\in T(\C)$, if $x_j \in W$ and $x_i\in F_k$ then the value $\A_k^{\alpha_t(x_i)}(\alpha_t(x_j))$ equals $z[x_i][x_j]$ thanks to the conditions ($\dagger$) and ($\ddagger$). \hfill\break

	\noindent{\color{lipicsGray}\sffamily Construction of the parity game.}
	Note that the size of $\C$ is at most $\mathcal{O}(|\max_{1\leq i \leq n} |\A_i|)^{\mathcal{O}(m^2+n)}$. 
	In particular, when $|\varPsi|$ is fixed (i.e., $n$, $m$ and $|\varphi|$ are fixed), $\C$ can be constructed in polynomial time since the satisfiability of an existential Presburger formula with a fixed number of quantifiers is in \PTime~\cite{ac165798-766d-3e3a-a2ea-7fe0e314d50a}.
	To handle the quantifiers of $\varPsi$, we construct a parity tree automaton $\C'$ that do not take inputs. 
	Essentially, $\C'$ is constructed from $\C$ by releasing the existentially quantified variables through projections, i.e., leaving the tree automaton a non-deterministic choice while relaxing a dimension of the input tree.
	Universal quantifiers $\forall x : \varPsi'$ are treated as $\lnot \exists x : \lnot\varPsi'$, where each negation $\lnot$ requires the computation of the complement of the current tree automaton, and then induce an exponential blow up of the computation time~\cite{DBLP:books/ems/21/Loding21}.
	The parity game is constructed in $\PTime$ from a tree automaton $\C'$.
	The game proceeds with the even player first choosing a transition in the tree automaton, and then the odd player choosing a subtree.
	The even player wins iff the language of the tree automaton is not empty.
	Naturally, a resolver logic formula with only existential quantifiers do not require tree automata complementations.
	However, with a naive approach, a formula with only universal quantifiers may requires two complementations while none are necessary.
	This is because if the innermost quantifier is universal then the first complementation can be avoided by using $\lnot \varphi$ instead of $\varphi$ to construct $\C$.
	Additionally, if the outermost quantifier is universal then the last complementation can be avoided by constructing a parity game that is winning for the player with even objective if and only if the current tree automaton is empty.
	This is doable in $\PTime$ as before, but the players are swapped and the priority are increased by one.
\end{proof}

\subsubsection*{Proofs of \cref{sec:applications}}

\begin{proof}[Proof of \cref{cl:bottom}]
	If $\A$ has a non-integral weight, simply multiply all weights with the appropriate constant.
	The bottom value of $\A$ is $x$ iff $\inf_{w \in \Sigma^\omega} \A_{\sup}(w) = x$.
	The first conjunct $\A^{f_1}(w_1) = x$ requires that there exists a word $w_1$ and a resolver $f_1$ such that the corresponding run of $\A$ yields the value $x$.
	For the bottom value of any $\Inf$-, $\Sup$-, $\LimInf$-, or $\LimSup$- automata, this holds since there are finitely many transitions and thus weights.
	The second conjunct $\A^{f_1}(w_1) \geq \A^{f_2}(w_1)$ requires that there is no other run of $\A$ over $w_1$ to exceed the value of this run.
	This condition forces resolving nondeterminism by supremum as usual for quantitative automata.
	Note that, by \cref{cl:supsupsup}, there exists a resolver to yield the supremum value.
	The third conjunct $\A^{f_1}(w_1) \leq  \A^{f_3}(w_2)$ requires that for every word $w_2$ there is a run of $\A$ over $w_2$ whose value is at least the value of the run of $\A$ produced by $f_1$ and $w_1$.
	Finally, this condition guarantees that no word has an $\A_{\sup}$ smaller than $w_1$, therefore $\A$ yields its bottom value on $w_1$.
	
	Combining the first two conjuncts together, we obtain that $\A$ has a run $\pi$ on some word $w$ whose value is $x$ and there is no other run on $w$ to produce a value greater than that of $\pi$, i.e., the value of $\pi$ is $\A_{\sup}(w) = x$.
	Finally, the third conjunct gives us that there is no other word $w'$ with $\A_{\sup}(w') < \A_{\sup}(w) = x$, i.e., $\inf_{w'' \in \Sigma^\omega} \A_{\sup}(w'') = x$.
\end{proof}


\begin{proof}[Proof of \cref{cl:hdness}]
	Thanks to \cref{rm:hdness} we only need to show that $\A \blindsim \A$ iff there exists $f \in R(\A)$ such that for all $w \in \Sigma^\omega$ we have $\A_{\sup}(w) \leq \A^f(w)$.
	The right-hand side is equivalent to the following: there exists $f \in R(\A)$ such that for all $w \in \Sigma^\omega$ we have $\sup_{g \in R(\A)} \A^g(w) \leq \A^f(w)$.
	Moreover, this holds iff there exists $f \in R(\A)$ such that for all $g \in R(\A)$ and all $w \in \Sigma^\omega$ we have $\A^g(w) \leq \A^f(w)$, which coincides with the definition of $\A \blindsim \A$.
\end{proof}


\begin{proof}[Proof of \cref{cl:hyperpositive}]
	Let $H_{\A} = \{\A^f \st f \in R(\A)\}$ be the hyperproperty defined by $\A$.
	We only show $H_{\A} = \textsf{SP}$, and $H_{\B} =\textsf{CP}$ can be proved similarly.
	
	$H_{\A} \subseteq \textsf{SP}$:
	Let $f \in R(\A)$ be an arbitrary resolver.
	We want to show that $\A^f$ is a safety property, i.e., for all $w \notin \A^f$ there exists a prefix $u \prec w$ such that $uw' \notin \A^f$ for all $w' \in \Sigma^\omega$.
	If $\A^f = \Sigma^\omega$, we are done.
	Otherwise, let $w = \sigma_1 \sigma_2 \ldots$ be such that $w \notin \A^f$.
	Then, there exists $n \geq 1$ such that $f(s_0 \sigma_1 \ldots s_0, \sigma_n) = s_1$.
	Observe that after this point all infinite runs stay in $s_1$, and thus are rejecting.
	Therefore, for $u = \sigma_1 \ldots \sigma_n$, we have $uw' \notin \A^f$ for all $w' \in \Sigma^\omega$, and thus $\A^f$ is safe.
	
	$\textsf{SP} \subseteq H_{\A}$:
	Let $P \in \textsf{SP}$ be a safety property.
	We want to show that there exists $f \in R(\A)$ such that $P = \A^f$.
	Let $S = \{ u \in \Sigma^* \st \forall w' \in \Sigma^\omega : uw' \notin P\}$ be the set of bad prefixes of $P$.
	We define $f$ such that for every $u = \sigma_1 \ldots \sigma_n \in \Sigma^*$ and $\sigma \in \Sigma$ we have if $u \notin S$ and $u \sigma \in S$ then $f(s_0 \sigma_1 \ldots \sigma_n s_0, \sigma) = s_1$, if $u \notin S$ and $u \sigma \notin S$ then $f(s_0 \sigma_1 \ldots \sigma_n s_0, \sigma) = s_0$.
	Note that $\A$ is deterministic otherwise.
	Then, for all $w \in \Sigma^\omega$, we have $w \notin \A^f$ iff there exists a prefix $u \prec w$ such that $u \in S$ iff $w \notin P$.
\end{proof}

\begin{proof}[Proof of \cref{cl:hypernegative}]
	We show that $H_1$ is not expressible by an automaton (the proof for $H_2$ is similar).
	For this, we prove that for every automaton $\A$ if $H_1 \subseteq H_{\A} = \{\A^f \st f \in R(\A)\}$ then $H_1 \subsetneq H_{\A}$.
	
	Observe that each $P \in H_1$ is characterized by a set $S \subseteq \{1, 2, \ldots\}$ of indices such that all $\sigma_1 \sigma_2 \ldots \in P$ and all $i \in S$ satisfy $\sigma_i = b$.
	
	Let $\A$ be an automaton and assume $H_1 \subseteq H_{\A}$.
	In particular, there exist distinct $P_1, P_2 \in H_1$ such that their characterizing sets $S_1, S_2 \subseteq \{1, 2, \ldots\}$ of indices do not contain 1, i.e., all words $w \in S_1 \cup S_2$ have either $a$ or $c$ as the first letter.
	By our initial assumption, there exist $f_1, f_2 \in R(\A)$ such that $\A^{f_1} = P_1$ and $\A^{f_2} = P_2$.
	
	Now, we describe a resolver $f \in R(\A)$ such that $\A^f \notin H_1$.
	Intuitively, $f$ imitates either $f_1$ or $f_2$ depending on the first letter of the input word, and since $\A^{f_1} \neq \A^{f_2}$, the trace property $\A^f$ contains two words on which the $b$-indices do not match.
	
	We define $f$ such that, given an arbitrary history $h = q_0 \sigma_1 q_1 \sigma_2 \ldots$ of $\A$ and a letter $\sigma \in \Sigma$, we have $f(h, \sigma) = f_1(h, \sigma)$ if $\sigma_1 = a$, and $f(h, \sigma) = f_2(h, \sigma)$ otherwise.
	Consider two words $w_1 \in P_1$ and $w_2 \in P_2$ such that $w_1$ starts with $a$ and $w_2$ starts with $c$.
	Then, with input $w_1$ the resolver $f$ produces the same run of $\A$ as $f_1$, which leads to acceptance of $w_1$, i.e., $w_1 \in \A^f$.
	Moreover, with input $w_2$, the resolver $f$ imitates $f_2$, therefore $w_2 \in \A^f$ as well.
	However, since $P_1$ and $P_2$ are distinct, the occurrences of $b$s in $w_1$ and $w_2$ do not overlap, therefore there is no $P \in H_1$ such that $u_1, u_2 \in P$, which implies $\A^f \not \in H_1$.
\end{proof}

\begin{proof}[Proof of \cref{cl:hyperinclusion}]
	Recall that, for a given boolean automaton $\A$, we have $H_{\A} = \{\A^f \st f \in R(\A)\}$ where each $\A^f \subseteq \Sigma^\omega$ is a trace property.
	Evidently, $H_{\A} \subseteq H_{\B}$ iff for every element of $H_{\A}$ also belongs to $H_{\B}$, i.e., for each $f \in R(\A)$ there exists $g \in R(\B)$ such that $\A^f = \B^g$, or equivalently $\A^{f}(w) = \B^{g}(w)$ for all $w \in \Sigma^\omega$.
\end{proof}
	
\end{document}